\documentclass[twocolumn]{aastex631}

\usepackage{color}
\usepackage{amsmath}

\newcommand{\rev}[1]{\textcolor{black}{#1}}

\begin{document}
\title{Transients by Black Hole Formation from Red Supergiants: Impact of Dense Circumstellar Matter}
\author[0000-0002-6347-3089]{Daichi Tsuna}
\affiliation{TAPIR, Mailcode 350-17, California Institute of Technology, Pasadena, CA 91125, USA}
\affiliation{Research Center for the Early Universe (RESCEU), School of Science, The University of Tokyo, 7-3-1 Hongo, Bunkyo-ku, Tokyo 113-0033, Japan}

\author[0000-0003-2868-489X]{Xiaoshan Huang}
\affiliation{TAPIR, Mailcode 350-17, California Institute of Technology, Pasadena, CA 91125, USA}

\author[0000-0002-4544-0750]{Jim Fuller}
\affiliation{TAPIR, Mailcode 350-17, California Institute of Technology, Pasadena, CA 91125, USA}

\author[0000-0001-6806-0673]{Anthony L. Piro}
\affiliation{Carnegie Observatories, 813 Santa Barbara Street, Pasadena, CA 91101, USA}

\correspondingauthor{Daichi Tsuna}
\email{tsuna@caltech.edu}

\begin{abstract}
Failed supernovae (SNe), which are likely the main channel for forming stellar-mass black holes, are predicted to accompany mass ejections much weaker than typical core-collapse SNe. We conduct a grid of one-dimensional radiation hydrodynamical simulations to explore the emission of failed SNe from red supergiant progenitors, leveraging recent understanding of the weak explosion and the dense circumstellar matter (CSM) surrounding these stars. We find from these simulations and semi-analytical modeling that diffusion in the CSM prolongs the early emission powered by shock breakout/cooling. The early emission has peak luminosities of $\sim 10^7$--$10^8~L_\odot$ in optical and UV, and durations of days to weeks. The presence of dense CSM aids detection of the early bright peak from these events via near-future wide-field surveys such as Rubin Observatory, ULTRASAT and UVEX.
\end{abstract}

\section{Introduction}
The origin(s) of stellar-mass black holes (BHs) remains an unsolved problem, with open questions on how BHs are born, and from what kind of stars \citep[e.g.,][]{OConnor11,Ugliano12,Kochanek14,Clausen15,Pejcha15,Sukhbold16,Ertl16,Muller16,Burrows20,Wang22,Boccioli24}. Ultimately, solving these requires observing BHs at their births. While failed supernovae (SNe) are believed to be the main channel to form BHs from massive stars, direct observations of them are still scarce. A survey monitoring nearby supergiants found candidates of vanishing stars without bright SNe \citep[e.g.,][]{Gerke15,Reynolds15,Neustadt21}, whose nature is still under debate \citep{Kashi17,Adams17a,Beasor24,Kochanek24}.

Another possibility explored in this work is to capture failed SNe as brightening (instead of vanishing) events. Before implosion to a BH, the collapsing core generally undergoes a protoneutron star phase emitting copious neutrinos. The neutrino emission reduces the gravitational mass of the core, which disrupts the hydrostatic equilibrium of the envelope and may trigger a weak explosion with energies of $10^{46}$--$10^{47}$ erg \citep{Nadyozhin80,Lovegrove13,Piro13,Fernandez18,Coughlin18a,Tsuna20, Ivanov21,daSilva23}.

Red supergiants (RSGs), the most common endpoints of massive stars, have inflated convective envelopes with low binding energy ($10^{47}$--$10^{48}$ erg). In the likely case where the inner part of the envelope cannot be ejected by the aforementioned neutrino mechanism, the angular momentum originating from non-radial convective motions in the envelope allows the infalling material to circularize outside the newborn BH \citep{Gilkis14,Quataert19,Antoni22}. This subsequently powers outflows that reverse the infall and unbind most of the envelope, with final energies ($\gtrsim 10^{48}$ erg) greatly exceeding the aforementioned neutrino mechanism \citep{Antoni23}.

These transients have evaded firm detection, likely due to their low luminosities compared to normal SNe. However, transient surveys are rapidly evolving, with near-future optical/UV surveys like Rubin Observatory \citep{Ivezic19}, ULTRASAT \citep{Ben-Ami22} and UVEX \citep{Kulkarni21} expected to probe into the dynamic universe at unprecedented depths. Such surveys have the potential to uncover novel dim transients, including the weak explosions from failed SNe. To conclusively identify these transients, detailed predictions on their observational signatures are important.

A recent key discovery from high-cadence observations of Type II SNe is the common presence of dense confined circumstellar matter (CSM) in RSGs \citep[e.g.,][]{Yaron17,Morozova17,Morozova18,Forster18,Bruch21,Irani23,Jacobson-Galan24}. This CSM, with a typical extent of $\sim 10^{14}$--$10^{15}$ cm, mainly affects the early phase of the explosion around shock breakout. While light curve modeling exists for weak explosions like failed SNe \citep[][see also \citealt{Dessart10}]{Piro13,Lovegrove17}, a detailed investigation of the impact of dense CSM has not been performed.

\begin{figure*}
    \centering
    \begin{tabular}{cc}
     \begin{minipage}[t]{0.5\hsize}
    \centering
    \includegraphics[width=\linewidth]{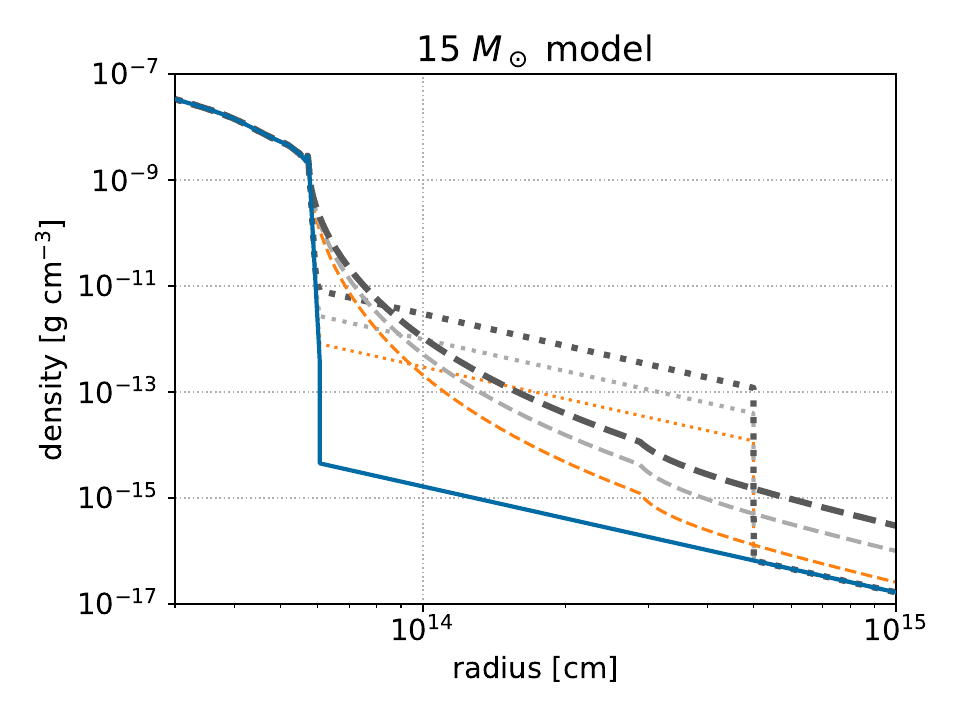}
    \end{minipage}
     \begin{minipage}[t]{0.5\hsize}
   \centering
    \includegraphics[width=\linewidth]{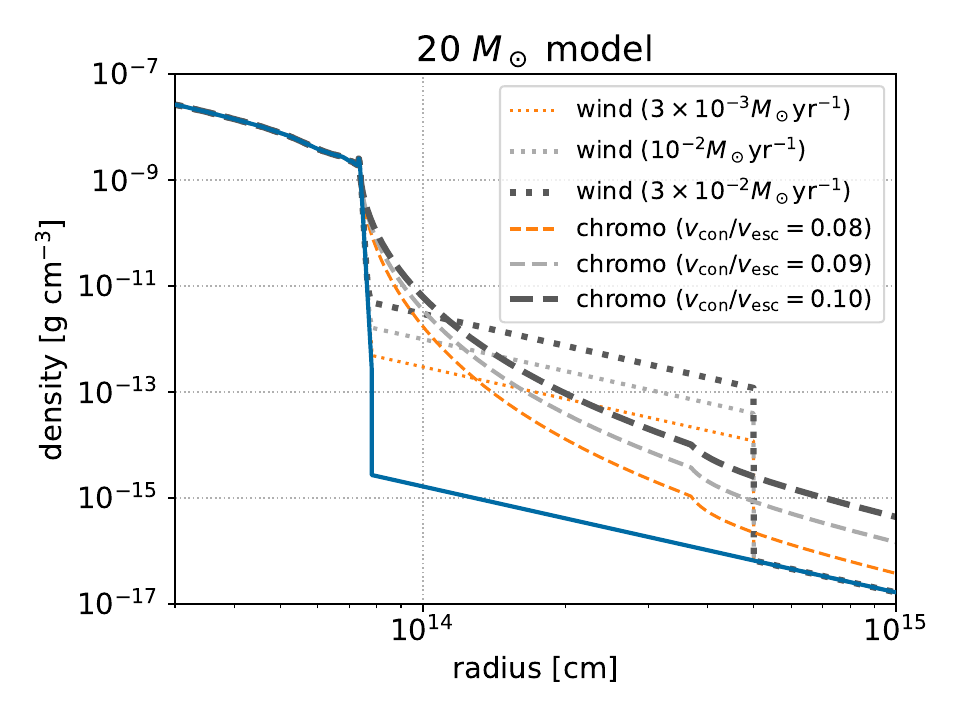}
    \end{minipage}
    \end{tabular}
    \caption{Density profiles of the CSM in our simulations, for the $15~M_\odot$ and $20~M_\odot$ models. The solid line shows the case of no dense CSM, that attaches a profile of a stellar wind with mass loss rate $\dot{M}= 10^{-5}~M_\odot$ yr$^{-1}$ and velocity $v_{\rm wind}=30$ km s$^{-1}$. Dashed lines are the ``dense chromosphere" CSM model of \cite{Fuller24}, and dotted lines are the ``superwind" model, with wind profiles of varying $\dot{M}$ inferred by \cite{Jacobson-Galan24}. Note that the inner boundary of our simulations is interior to the inner edge of this plot.}
    \label{fig:CSM_profile}
\end{figure*}

In this work we explore the observational signatures of failed SNe from RSGs using a grid of 42 one-dimensional radiation hydrodynamical simulations with varied progenitors, energy injection, and ambient CSM inspired from observations of Type II SNe. We find that like normal Type II SNe, the likely presence of similar CSM in failed SNe can impact the early light curve morphology. The early peak is prolonged due to the CSM, which aids detection by current and near future optical/UV surveys with cadences of days.

This paper is constructed as follows. In Section \ref{sec:model} we present the details of our simulations. In Section \ref{sec:results} we show the variations of resulting light curves for the parameters governing the explosion and the CSM. We discuss the implications for ongoing and near-future transient surveys in Section \ref{sec:discussion}, and conclude in Section \ref{sec:conclusion}.

\section{Model}
\label{sec:model}
We use the Lagrangian one-dimensional radiation hydrodynamical code SNEC \citep{Morozova15}, to model the mass ejections of failed SNe from RSGs and obtain their light curves. We describe our model in detail, with the RSG models in Section \ref{sec:RSG_models}, CSM models in Section \ref{sec:CSM_models}, parameterized explosion modeling in Section \ref{sec:energy_inj_models}, and the detailed setup of SNEC in Section \ref{sec:SNEC_setup}.

\subsection{RSG Prognenitor Models}
\label{sec:RSG_models}
Massive stars of zero-age main sequence (ZAMS) masses from $\approx 8~M_\odot$ and up to $20$--$30~M_\odot$ are believed to undergo core-collapse as RSGs \citep[e.g.,][]{Ekstrom12,Beasor20}, but the masses of stars that produce BHs is uncertain. In fact, the theoretical models generally do not predict a single well-defined mass range for failed SNe, but predict there are ``islands of explodability" \citep[e.g.,][]{Sukhbold16} whose locations are still under debate.

Given the uncertainties, we adopt non-rotating pre-SN RSGs with masses of $15~M_\odot$ and $20~M_\odot$ and solar metallicity at ZAMS, evolved until core carbon depletion with the code MESA \citep[r10398:][]{Paxton11,Paxton13,Paxton15,Paxton18,Paxton19,Jermyn23}. We use the \rev{inlist \citep{FT24_zenodo}} used in a recent work \citep{Fuller24}, which adopted a mixing-length parameter of $\alpha_{\rm MLT}=2.5$ and mass loss prescription developed in that work. We note that the derived mass loss rates at the core helium burning phase agree with empirical prescriptions within a factor of a few (Fig. 7 of \citealt{Fuller24}). The $15$, $20~M_\odot$ models have final masses $M_*$ of $13.9$, $16.3~M_\odot$, and helium core masses of $5.2$, $7.2~M_\odot$ respectively.

We then load and relax\footnote{We re-run the same inlist but with additional commands \texttt{relax\_to\_this\_tau\_factor=1.5d-5} and \texttt{dlogtau\_factor=0.1}.} the above MESA models for additional timesteps to include material outside the photosphere down to $\tau=10^{-5}$, in order for the density of the stellar models to more smoothly connect with that of the CSM. This leads to new RSG models with photospheric radii of a few percent different from the original ones, but otherwise very similar profiles. The new stellar profiles extend to densities as low as $\approx (3$--$4)\times 10^{-13}$ g cm$^{-3}$. The updated 15 and 20$~M_\odot$ models have photospheric radii $R_*$ of $825~R_\odot$ and $1059~R_\odot$ respectively.

\subsection{Circumstellar Matter Models}
\label{sec:CSM_models}
Observations of early light curves and spectra of Type II SNe point towards the existence of confined ($\lesssim 10^{15}$ cm), massive ($\sim 10^{-2}$--$1~M_\odot$) CSM in RSGs at core-collapse. Here we adopt two representative models for the dense CSM inspired from recent studies, as well as a comparison case without a dense CSM which were studied analytically and numerically in previous works \citep{Piro13,Lovegrove17,Fernandez18}. The density profiles of the CSM models are shown in Figure \ref{fig:CSM_profile}.

First, for the case without dense CSM we adopt a conventional (low-density) stellar wind of mass loss rate $\dot{M}=10^{-5}~M_\odot\ {\rm yr}^{-1}$ and velocity $v_{\rm wind}=30~$ km s$^{-1}$, with a wind profile $\rho=\dot{M}/(4\pi r^2 v_{\rm wind})\propto r^{-2}$. Such a wind has weak influence in the light curves, and the light curve is insensitive to the choices of $\dot{M}$ and $v_{\rm wind}$.

Next, for one of the scenarios with dense CSM we consider a case of $\dot{M}$ highly enhanced from a conventional wind (often dubbed as ``superwind"). We adopt a wind CSM profile with parameters inferred from model fitting for light curves and spectra of a large sample of young Type II SNe \citep[][]{Jacobson-Galan24}. Such observations probe the density profile (i.e. $\dot{M}/4\pi v_{\rm wind}$), but they do not strongly constrain $\dot{M}$ and $v_{\rm wind}$ individually. Here we consider three normalizations of $\dot{M}/v_{\rm wind}=[3\times 10^{-3}, 10^{-2},3\times 10^{-2}]~M_\odot$ yr$^{-1}/(50\ {\rm km\ s^{-1}})$ that cover the range of values inferred in \cite{Jacobson-Galan24} (their Fig. 15). We adopt a stationary CSM (i.e. $v_{\rm  CSM}=0$) to avoid the formation of artificial gaps between the star and the CSM\footnote{The assumption for CSM velocity $v_{\rm CSM}$ can become important for our failed SN models, as the shock velocity $v_{\rm sh}$ at the RSG surface can be as low as $\sim 100$ km s$^{-1}$ (Section \ref{sec:hydro_results}). The light curve is insensitive to $v_{\rm CSM}$ as long as $v_{\rm CSM}\ll v_{\rm sh}$ near the stellar surface, likely valid if the velocity profile is characterized by e.g., near-surface convection or an accelerating stellar wind.}. We set the outermost extent of the dense CSM to $r_{\rm out}=5\times 10^{14}$ cm \citep[Fig. 18 of][]{Jacobson-Galan24}, and attach the aforementioned low-density wind profile at $r>r_{\rm out}$.

Finally, we adopt a recent CSM model of \cite{Fuller24} that considers mass loss of RSGs, via a combined effect of (i) outgoing shock waves by near-surface convection and (ii) radiation pressure outside where dust forms. The model generally explains the observed mass-loss rates of local RSGs and their luminosity dependence, and predicts a confined dense ``chromosphere" with (angle/time-averaged) density profile
\begin{equation}
    \rho(r \leq R_{\rm d}) = \rho_*\left(\frac{R_*}{r}\right)^2\exp\left[-\left(\frac{v_{\rm esc}}{v_{\rm con}}\right)\sqrt{1-\frac{R_*}{r}}\right].
\end{equation}
Here $\rho_*$ is the density at $r=R_*$, $R_{\rm d}$ is the dust formation radius, and $v_{\rm esc}=\sqrt{2GM_*/R_*}$ is the surface escape velocity with $G$ being the gravitational constant. Note that the radial velocity of the shock waves at a given time is assumed to follow a Gaussian probability distribution with mean of the typical convective velocity $v_{\rm con}$. At $r>R_{\rm d}$, the density profile is given as
\begin{equation}
    \rho(r > R_{\rm d}) = \frac{\dot{M}}{4\pi r^2 v_{\rm out}(r)}
\end{equation}
with mass loss rate $\dot{M}$ set from the local values at the dust formation radius $\dot{M}=4\pi R_{\rm d}^2\rho(R_{\rm d})v_{\rm d}$, and velocity profile determined by radiation pressure on dust-laden material with Eddington ratio $f_{\rm Ed}(>1)$ as
\begin{eqnarray}
    v_{\rm out}(r>R_{\rm d})&=& \left[v_{\rm d}^2+2(f_{\rm Ed}-1)GM_*\left(\frac{1}{R_{\rm d}}-\frac{1}{r}\right)\right]^{1/2} \nonumber \\
    &\underset{r\to \infty}{\to}& \sqrt{v_{\rm d}^2+2(f_{\rm Ed}-1)\frac{GM_*}{R_{\rm d}}} \equiv v_{\rm \infty}.
\end{eqnarray}
Following \cite{Fuller24}, we set $v_{\rm d}\approx v_{\rm con}$, $R_{\rm d}=5R$, and the terminal wind velocity $v_{\infty}=30~$ km s$^{-1}$ (see their Section 3.1 for discussion on each parameter). \rev{The model has a finite velocity profile $v_{\rm CSM}(r)=\dot{M}/4\pi r^2 \rho(r)$ that is small ($\lesssim 10\ {\rm km\ s^{-1}}$) at $r<R_{\rm d}$ and asymptotes to $v_\infty$ at large radii (their Fig. 5), which we adopt in our simulations.}

The ratio $v_{\rm con}/v_{\rm esc}$ is the key parameter for the dense CSM, and is varied as $v_{\rm con}/v_{\rm esc}=[0.08, 0.09, 0.1]$. These values cover the expected range found in pre-SN RSG models, with $v_{\rm con}/v_{\rm esc}\approx 0.08$ (0.09) for our $15~M_\odot$ ($20~M_\odot)$ models. \cite{Fuller24} modeled the impact of this dense chromosphere on Type II SNe, and found $v_{\rm con}/v_{\rm esc}=0.1$ to roughly explain the early light curve of SN 2023ixf \citep{Jacobson-Galan23,Hiramatsu23,Li24,Zimmerman24} for an explosion energy of $10^{51}$ erg. 

A difference in the chromosphere model and the superwind model is the location of the bulk of the CSM mass. The former predicts most of the mass to be within $\sim 2R_*$ with a rapidly declining density profile, in contrast to a wind profile that has most of the mass near the outermost radii ($\sim 10~R_*$). We note that this model is not the only one to predict such a confined CSM, and massive CSM at $\lesssim$ (a few) $R_*$ have often been invoked to explain early light curves of Type II SNe \citep[e.g.,][but see also \citealt{Dessart23}]{Morozova18,Moriya18,Haynie21}.

\subsection{Energy Injection Modeling}
\label{sec:energy_inj_models}
\begin{figure*}
    \centering
    \begin{tabular}{cc}
     \begin{minipage}[t]{0.5\hsize}
    \centering
    \includegraphics[width=\linewidth]{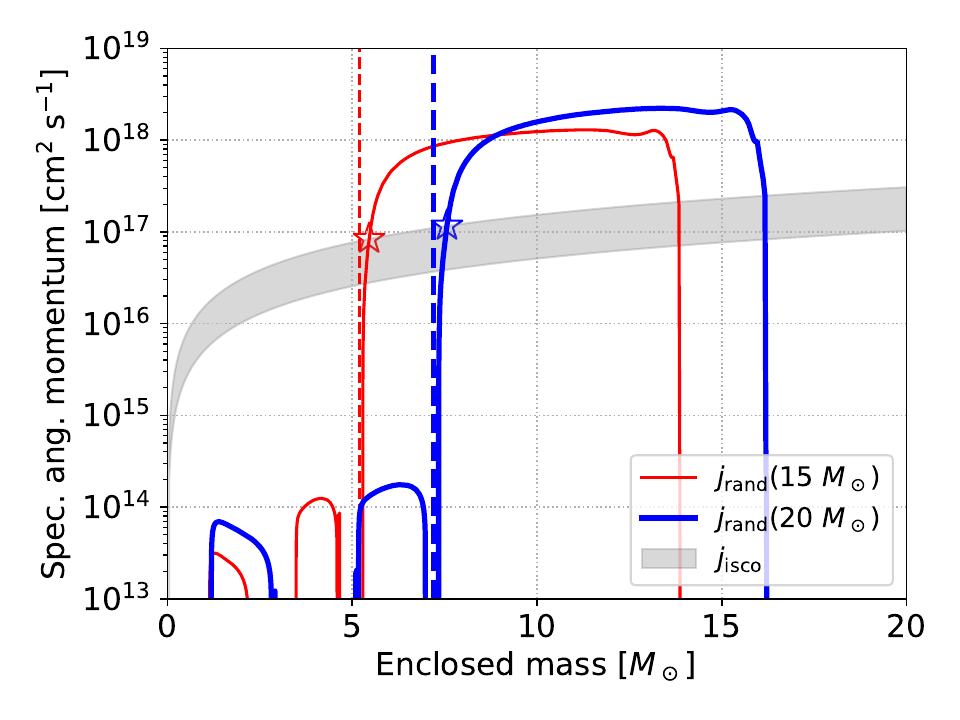}
    \end{minipage}
     \begin{minipage}[t]{0.5\hsize}
   \centering
    \includegraphics[width=\linewidth]{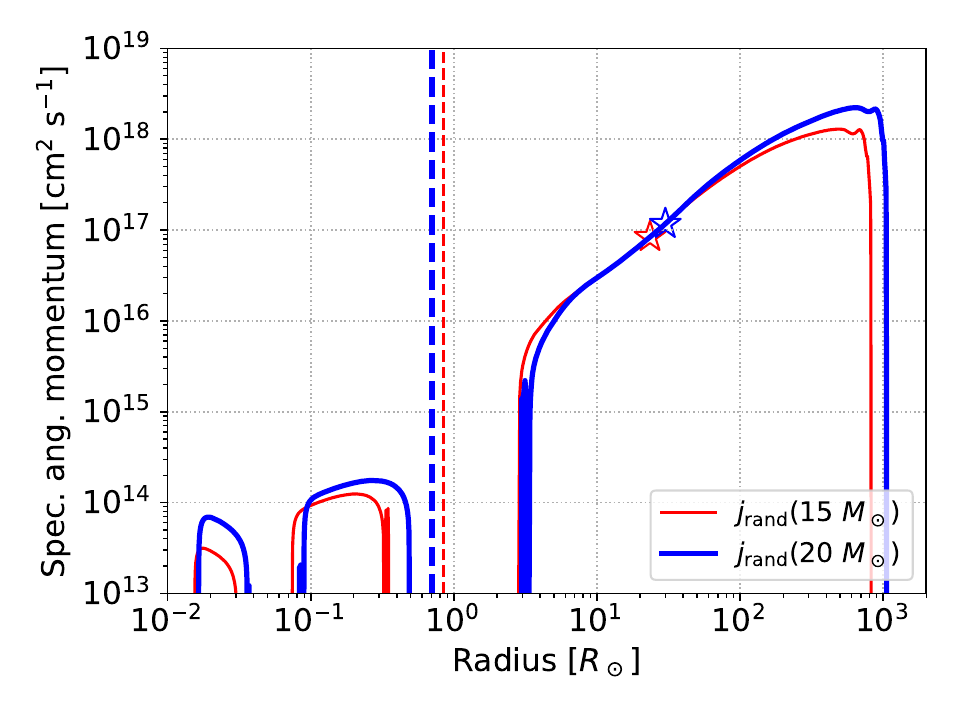}
    \end{minipage} \\
    \begin{minipage}[t]{0.5\hsize}
    \centering
    \includegraphics[width=\linewidth]{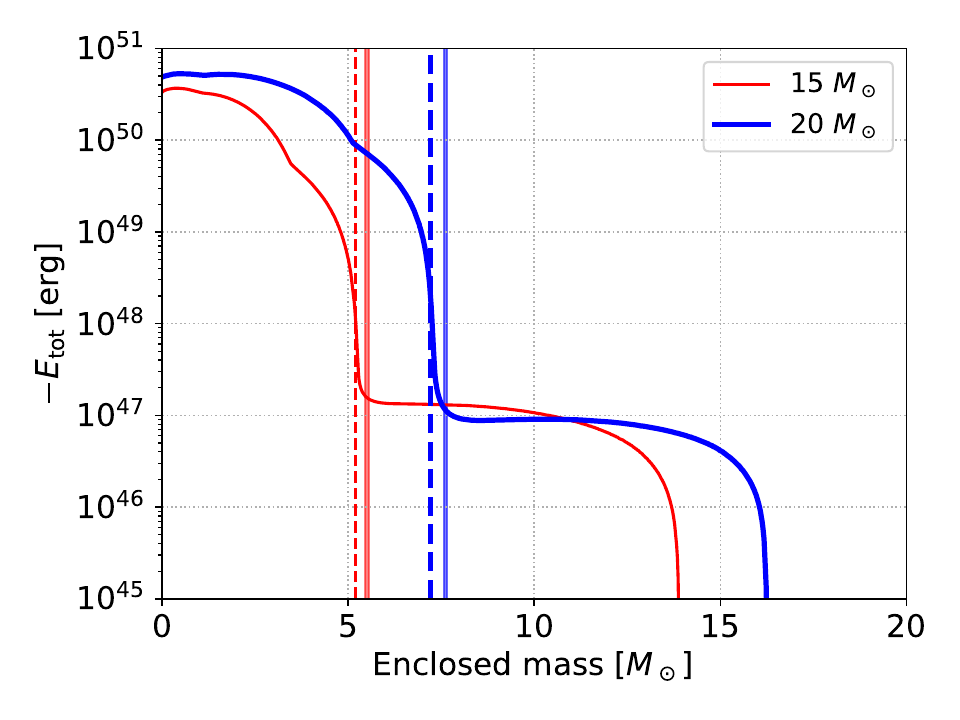}
    \end{minipage}
     \begin{minipage}[t]{0.5\hsize}
   \centering
    \includegraphics[width=\linewidth]{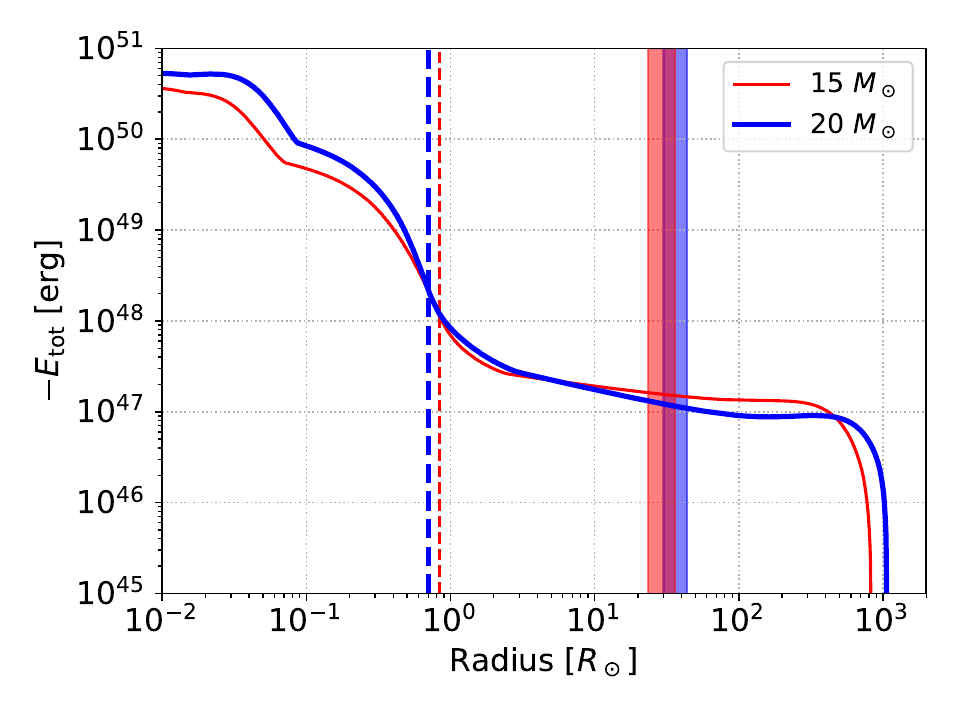}
    \end{minipage}
    \end{tabular}
    \caption{\rev{Specific angular momentum of random convective motion (top panels) and total energy exterior to a given mass/radius (bottom panels) in the RSG, as a function of mass (left panels) and radius (right panels).} Thin and thick lines are for models with initial masses of 15 and 20 $M_\odot$ respectively. The gray shaded region in the \rev{top left} panel shows the minimum angular momentum $j_{\rm ISCO}$ required for material to circularize outside the ISCO of a BH of given mass in the horizontal axis, with uncertainty reflecting the BH's spin (higher $j_{\rm ISCO}$ for lower spins). Stars denote the locations we set as inner boundaries of the SNEC simulation, where $j_{\rm rand}=j_{\rm ISCO}$ for a Schwarzschild BH. Vertical dashed lines show the location of the helium core\rev{, and vertical shaded regions in the bottom panels show regions where energy is injected in the SNEC simulations}.}
    \label{fig:jrand}
\end{figure*}

The mechanism(s) that triggers the ``explosion" in failed SNe completely differs from normal SNe. Hence the desired energy, location and timescale of the injection for failed SNe will all be very different from normal SNe, where a $\sim 10^{51}$ erg explosion is driven in the innermost core within an order of seconds. Accurately predicting these quantities is difficult, for both scenarios of ejection by core neutrino mass loss and by non-radial infall of the convective envelope. 

The latter scenario has recently been investigated in \cite{Antoni22,Antoni23}, via three-dimensional hydrodynamical simulations of envelope convection and infall of the hydrogen-rich envelope onto the newborn BH. The energy of the final explosion was found to be $\gtrsim 10^{48}$ erg, which overwhelms the range of energies recently estimated for the core neutrino mass loss scenario ($10^{46}$--$3\times 10^{47}$ erg; \citealt{Ivanov21,daSilva23}). Moreover this outflow energy is likely a lower limit, due to limitations in the inner boundary of their simulations being $\sim$100 times outside the BH's innermost stable circular orbit (ISCO). Based on scalings found in supercritical accretion flows onto BHs, they suggest that the energy injection could be up to $\sim$10 times higher if the accretion were to be simulated down to the ISCO.

We thus model the explosion under the latter envelope circularization scenario, but given its uncertainties we take a parameterized approach as follows. Using our MESA models we obtain the random angular momentum profile in the star as \citep{Quataert19}
\begin{eqnarray}
    j_{\rm rand} &\approx& \frac{H_{\rm p}(r)v_{\rm conv}(r)}{\sqrt{4\pi}} \nonumber \\
                    &\sim& 2\times 10^{18}\ {\rm cm^2\ s^{-1}}\left(\frac{H_{\rm p}}{100~R_\odot}\right)\left(\frac{v_{\rm conv}}{10~{\rm km\ s^{-1}}}\right)
\end{eqnarray}
where $H_{\rm p}(r)$ and $v_{\rm conv}(r)$ are the local pressure scale height and the convective velocity. We compare $j_{\rm rand}$ with the required angular momentum for gas to circularize outside the ISCO of the BH
\begin{eqnarray}
    j_{\rm ISCO} &=& \frac{f_{a}(2\sqrt{3}GM_{\rm BH})}{c} \nonumber \\
                &\sim& 8\times 10^{16}\ {\rm cm^2\ s^{-1}} \left(\frac{f_{a}M_{\rm BH}}{5~M_\odot}\right)
\end{eqnarray}
where $M_{\rm BH}$ is the BH's mass, $c$ is the speed of light, and $f_{a}$ is a factor that depends on the BH's spin with $f_{a}=1$ ($1/3$) for a Schwartzschild (maximum Kerr) BH.
Figure \ref{fig:jrand} plots $j_{\rm rand}$ as a function of mass and radius coordinates, or enclosed mass for $j_{\rm ISCO}$. Gas outside the helium core (indicated by vertical dashed lines) carries sufficient angular momentum to circularize outside the BH's ISCO, while gas interior would fall almost radially and be swallowed by the newborn BH.

We find the location $r_{\rm crit}$ that satisfies $j_{\rm rand}(r_{\rm crit})=j_{\rm ISCO}$, where the $M_{\rm BH}$ in $j_{\rm ISCO}$ corresponds to mass interior to $r_{\rm crit}$, and we set $f_{a_*}=1$ as the BH is expected to initially be almost non-rotating \citep[e.g.,][]{Fuller_Ma_19}. We excise the material interior to this mass, assuming this has fallen into the BH before the energy injection. The radii $r_{\rm crit}$ ($\approx$ a few $10~R_\odot$) and mass coordinates of this inner boundary for the $15$ and $20~M_\odot$ models are indicated as star symbols in Figure \ref{fig:jrand}.

\begin{table*}[]
    \centering
    \begin{tabular}{cc}
         \hline
         Parameters &  Values\\
         \hline
         Progenitor ZAMS mass [$M_\odot$] & $15$, $20$\\ \hline
         & no dense CSM ($\dot{M}=10^{-5}~M_\odot$ yr$^{-1}$) \\
         Dense CSM profile & superwind ($\dot{M}=3\times 10^{-3},10^{-2},3\times 10^{-2}~M_\odot\ {\rm yr}^{-1}$) \\
         & dense chromosphere ($v_{\rm con}/v_{\rm esc}=0.08, 0.09, 0.1$) \\ \hline
         Final explosion energy ($E_{\rm exp}$ [erg]) & $10^{48}$, $3\times 10^{48}$, $10^{49}$  \\
         Energy injection mass coordinate ($[M_\odot]$; fixed) & $\approx 5.46$ ($15~M_\odot$ model),  $\approx  7.56$ ($20~M_\odot$ model)\\
         Energy injection duration (days; fixed) & $\approx 0.898$ ($15~M_\odot$ model),  $\approx  1.10$ ($20~M_\odot$ model)
    \end{tabular}
    \caption{Variation of models carried out in this study. See main text for details of each model.}
    \label{tab:params}
\end{table*}

We then explode the stellar models (envelope + CSM) by the thermal bomb setup in SNEC. The explosion energy is uncertain as discussed above, with an expected range of $10^{48}$--$10^{49}$ erg. We vary the energy as a free parameter with three values of $E_{\rm exp}=[10^{48}, 3\times 10^{48}, 10^{49}]$ erg. Note that this is the final explosion energy taking into account the (negative) total energy $E_{\rm tot}$ of the computational region. The actual energy injected to the envelope is larger than $E_{\rm exp}$ by the absolute value of the total energy, which is $|E_{\rm tot}|\approx$ $1.4 \times 10^{47}$ erg ($9.5 \times 10^{46}$ erg) for the $15~M_\odot$ ($20~M_\odot$) model. \rev{The energy injection is done over the innermost $0.1~M_\odot$ of the stellar model.}

The duration of the thermal bomb $\Delta t_{\rm bomb}$ is set as the free-fall timescale at $r_{\rm crit}$, 
\begin{eqnarray}
    \Delta t_{\rm bomb} &\approx& \sqrt{\frac{r_{\rm crit}^3}{GM_{\rm BH}}} \nonumber \\
    &\sim& 1~{\rm day}\left(\frac{r_{\rm crit}}{25~R_\odot}\right)^{\!\!3/2}\left(\frac{M_{\rm BH}}{5~M_\odot}\right)^{\!\!-1/2}.
\end{eqnarray}
This assumes that the timescale and energy budget is dominated by the infall of gas at around $r_{\rm crit}$, as this can circularize closest to the BH and hence the resulting outflow likely carries the largest amount of energy. This is qualitatively found in the simulations of \cite{Antoni23}, where employing a smaller inner boundary (i.e. allowing more material to circularize closer to the BH) leads to faster and larger energy injection (their Figure 12). 

We note that setting $r_{\rm crit}$ is subtle, due to limitations on approximating $j_{\rm rand}(r)$ by the local variables as done in Figure \ref{fig:jrand}. Simulations of RSG convection \citep[e.g.,][]{Antoni22, Goldberg22a} find the convection to be more global, and the slope of $j_{\rm rand}(r)$ is shallower than the local approximation as the inner region inherits the larger $j_{\rm rand}$ from the outer layers. Furthermore, in reality $j_{\rm rand}(r)$ is not a single value but has a distribution, and some fraction of material at radii $r<r_{\rm crit}$ may also be able to circularize despite the average $j_{\rm rand}$ being less than $j_{\rm ISCO}$. These effects would both lead to an effectively shorter $\Delta t_{\rm bomb}$, but for a fixed $E_{\rm exp}$ we expect the results are insensitive to $\Delta t_{\rm bomb}$ as long as it is much shorter than the dynamical timescale of the envelope $t_{\rm dyn}\approx \sqrt{R_*^3/GM_*} \sim 100~{\rm day}\ (R_*/800~R_\odot)^{3/2}(M_*/{15~M_\odot})^{-1/2}$ \citep[see also Appendix of][]{Tsuna23}.

\subsection{Computational Setup in SNEC}
\label{sec:SNEC_setup}

In summary, we attach the parameterized CSM models to the $15~M_\odot$ and $20~M_\odot$ MESA RSG models as done in Figure \ref{fig:CSM_profile}. We excise the inner regions which are expected to be simply swallowed by the newly formed BH (inside the star symbols in Figure \ref{fig:jrand}). The models are then exploded by a thermal bomb with parameterized explosion energies, with a timescale of $\approx 1$ day reflecting the infall of the innermost region of the envelope capable of circularizing outside the BH. The parameters for the CSM and explosion are summarized in Table \ref{tab:params}.

For all CSM models we assume abundances equivalent to the surface abundances of the MESA models, and an initial temperature profile of a spherical, optically thin chromosphere under radiative equilibrium 
\begin{eqnarray}
    T_{\rm CSM}(r\geq R_*) = T_* \left[\frac{1-\sqrt{1-\left(R_*/r\right)^2}}{2}\right]^{1/4},
\end{eqnarray}
where $T_*$ is the photospheric temperature of the MESA model. The profile and abundance at $r<R_*$ are set to be the same as the MESA model, but variations in the abundance within the computational region are tiny (within $0.1\%$).

We adopt a grid of 3000 Lagrangian cells, enhanced from the default gridding of 1000 cells in SNEC, to adequately resolve the shock breakout region (near the stellar surface) as well as the lower density CSM. The initial outer boundary of the CSM is set to $10^{15}$ cm.

\begin{figure}
   \centering
    \includegraphics[width=\linewidth]{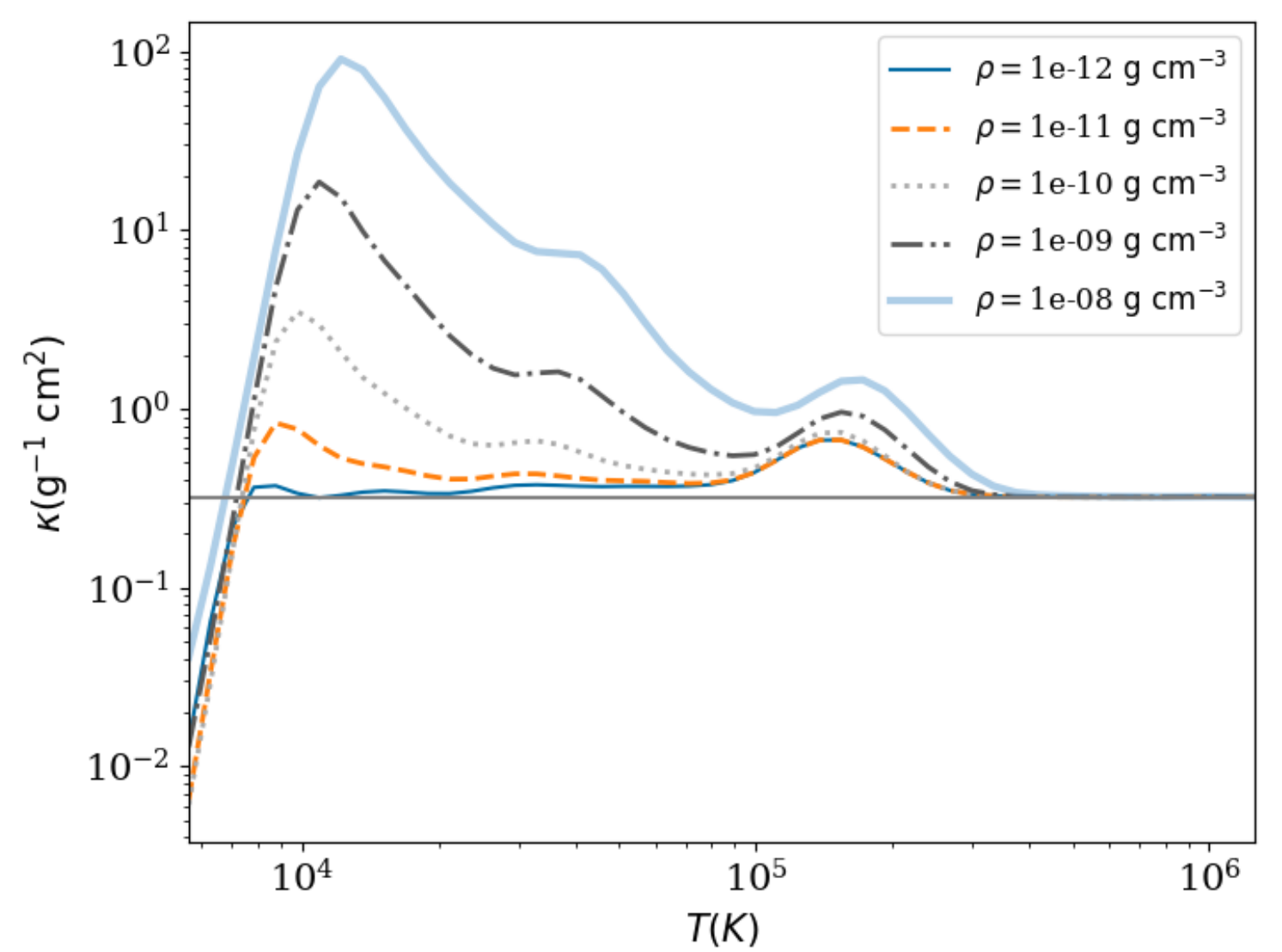}
    \caption{OPAL Rosseland mean opacities used in SNEC, for solar metallicity and density range of $10^{-12}$--$10^{-8}$ g cm$^{-3}$ relevant for our work. The horizontal line shows the Thomson scattering opacity for fully ionized gas.}
    \label{fig:kappa_OPAL}
\end{figure}

SNEC uses the Rosseland mean opacity tables of OPAL \citep{Iglesias96} and low-temperature tables by \cite{Ferguson05} as input. As a demonstration, the OPAL opacities for densities most relevant to this work of $\rho =10^{-12}$--$10^{-8}$ g cm$^{-3}$ with solar metalicity are plotted in Figure \ref{fig:kappa_OPAL}. Thomson scattering is a good approximation for the total opacity only at low densities of $\lesssim 10^{-11}~{\rm g\ cm^{-3}}$ or high temperatures of $\gtrsim$ (a few) $\times 10^5$ K, and other opacities (H$^{-}$, bound-free) contribute at higher densities and/or lower temperatures. In contrast to normal SN II-P, such deviations from Thomson scattering are important for characterizing the light curves of weak explosions expected in failed SNe.

%For normal Type II-P SNe electron scattering is reasonable, as low densities are achieved in the plateau phase at $\gtrsim 1$ month from explosion \dt{(NOTE: $10~M_\odot/[4\pi ({\rm month} \times 3000\ {\rm km\ s^{-1}})^3/3]\sim 10^{-11} $ g/cm$^3$)}. The density is retained much higher in these low-energy explosions such that the opacity is much higher.

\section{Results}
\label{sec:results}

\begin{figure*}
   \centering
    \begin{tabular}{cc}
     \begin{minipage}[t]{0.5\hsize}
    \centering
    \includegraphics[width=.95\linewidth]{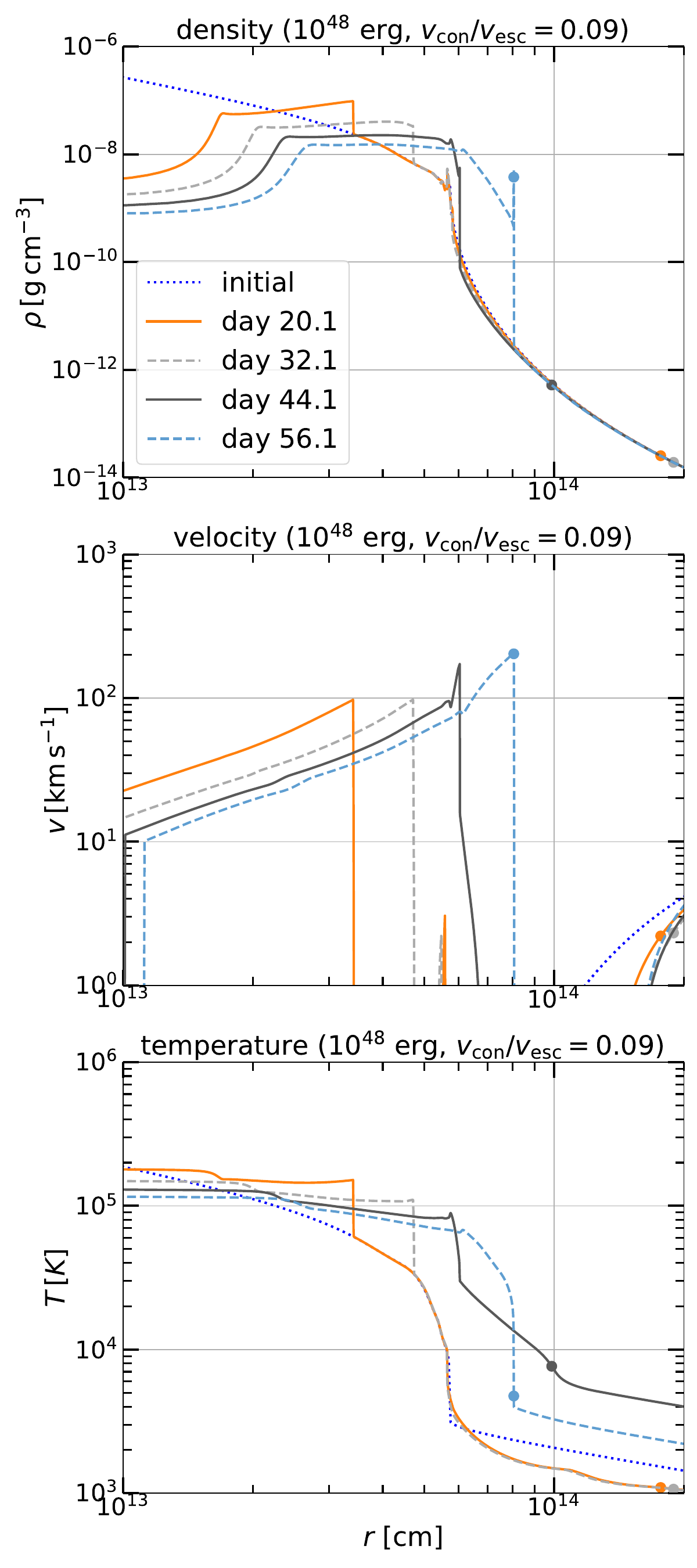}
    \end{minipage}
     \begin{minipage}[t]{0.5\hsize}
   \centering
    \includegraphics[width=.95\linewidth]{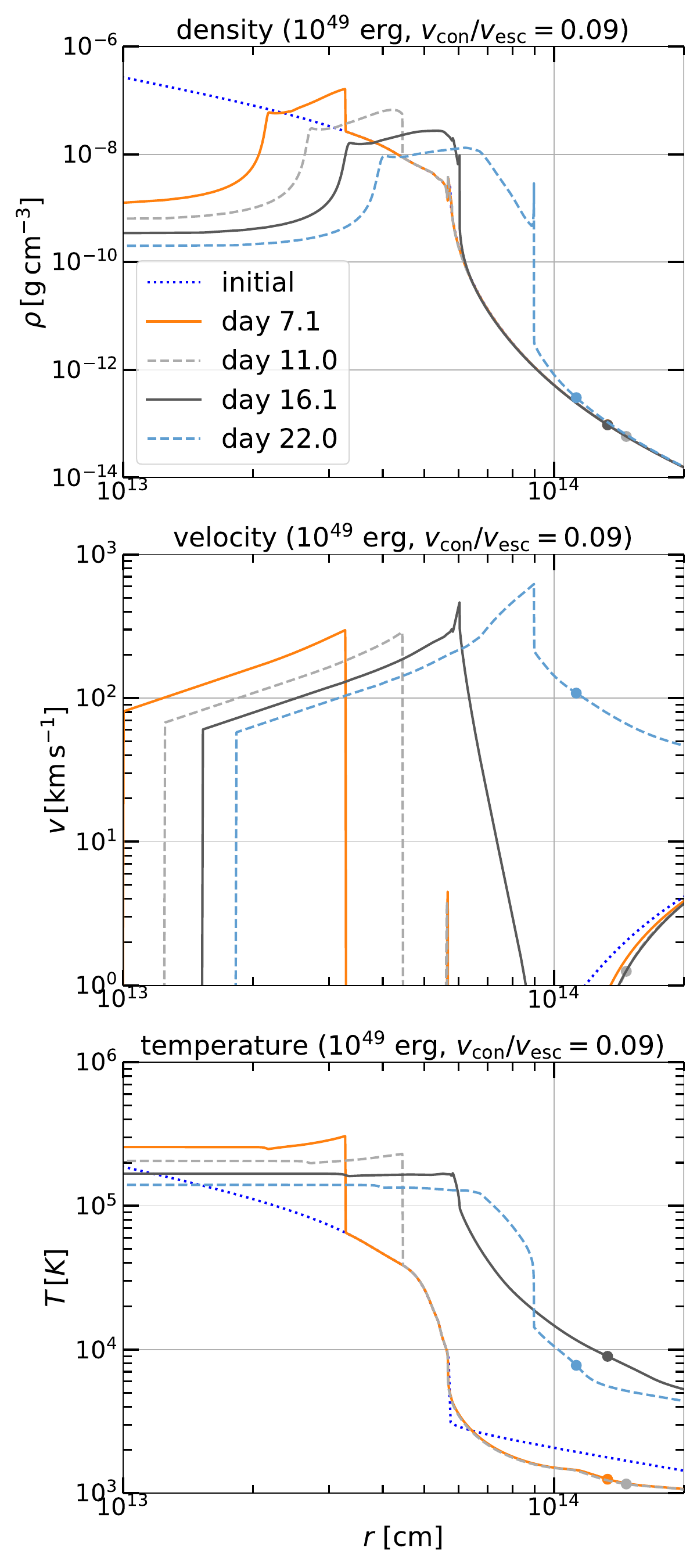}
    \end{minipage} 
    \end{tabular}
    \caption{Density, velocity and temperature profiles simulated by SNEC at around shock breakout, for explosions of the $15~M_\odot$ RSG model with energies of $10^{48}$ erg (left panels) and $10^{49}$ erg (right panels). We plot the case of the chromosphere CSM model with $v_{\rm con}/v_{\rm esc}=0.09$. The blue dotted lines show the initial profiles. The dots show the location of the photosphere at each epoch. Note that we plot profiles at different epochs for different explosion energies.}
    \label{fig:rho_v_T_profiles}
\end{figure*}

The results of the hydrodynamical evolution of the shock and the expanding ejecta are presented in Section \ref{sec:hydro_results}, including bolometric light curves. In Section \ref{sec:lc_noCSM} we first present the cases for no dense CSM also studied in previous works \citep{Piro13,Lovegrove13,Lovegrove17}, and then investigate the effect of dense CSM at the early phase of the light curve in Section \ref{sec:lc_withCSM}. 

\subsection{Hydrodynamical Evolution}
\label{sec:hydro_results}

Figure \ref{fig:rho_v_T_profiles} shows the hydrodynamical profiles calculated by SNEC with explosion energies of $E_{\rm exp}=10^{48}, 10^{49}$ erg, for selected snapshots around shock breakout. To illustrate the effect of the breakout on the ambient CSM, we adopt a model with dense CSM, with the chromosphere model of $v_{\rm con}/v_{\rm esc}=0.09$ as a representative case.

Soon after energy injection a shock develops and propagates through the star, similar to normal core-collapse SNe but with much slower velocity ($v_{\rm sh}\approx 100$--$300$ km s$^{-1}$ for these energies). As the shock reaches the surface, it accelerates slightly due to the steeper density gradient at the surface and the ambient CSM. Shock breakout happens at the stellar surface, and after that radiation can efficiently escape to the upstream CSM. Much after breakout, a thin shell of radiatively cooled shocked CSM is developed, seen as a density spike in the top panels. A recombination front develops and propagates inside the ejecta, powering plateau emission as in Type II-P SNe. We find that most of the mass in the computational region ($>99\%$) is eventually unbound.

Upon breakout, the ambient CSM is heated by the breakout pulse, to temperatures of $\sim 10^4$ K where hydrogen becomes ionized and the opacity is enhanced. Photon diffusion in this optically thick CSM prolongs the breakout signal, as will be seen later in this section. The unshocked CSM cools soon after, but becomes heated again after it is swept by the shock.

In the $10^{49}$ erg model we see a significant increase in the velocity of the upstream CSM to $\sim 100~{\rm km \ s^{-1}}$, which we attribute to acceleration by photons emitted in the breakout and later CSM interaction \citep{Chugai02,Tsuna23b}. The accelerated CSM may be seen as narrow components or systematic blueshifts in the early spectra. SNEC assumes local thermal equilibrium (LTE), which certainly breaks down in the lower-density CSM near the photosphere that is responsible for line emission. Non-LTE spectral modeling as done in Type II SNe \citep[e.g.,][]{Dessart17,Hillier19,Boian20} is required to make detailed predictions on line luminosities and their detectability.

\subsection{Models without Dense CSM}
\label{sec:lc_noCSM}
\begin{figure}
   \centering
    \includegraphics[width=\linewidth]{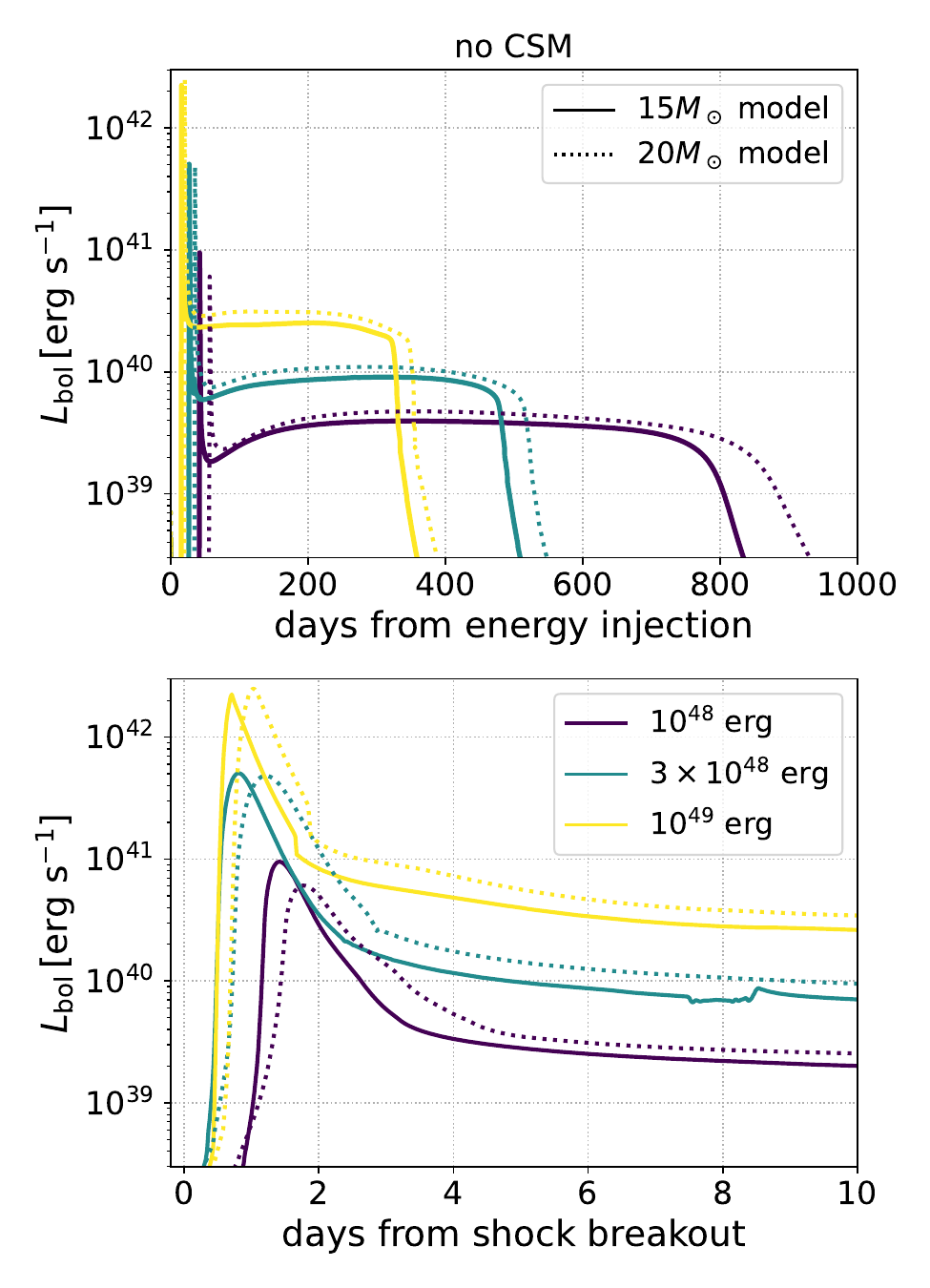}
    \caption{Bolometric light curves of explosions with no dense CSM, for final energies of $E_{\rm exp}=[10^{48}$, $3\times 10^{48}$, $10^{49}$] erg. Solid lines are for the $15~M_\odot$ model, and dotted lines are for the $20~M_\odot$ model. The bottom panel shows light curves from time of shock breakout (recorded in SNEC), for comparison of the early emission.}
    \label{fig:light_curves_noCSM}
\end{figure}

Figure \ref{fig:light_curves_noCSM} shows the light curves with no dense CSM, and varied explosion energy. For all explosion energies, the light curves display a brief shock breakout emission of order days, followed by a year-long plateau emission sustained by the recombining hydrogen envelope. These features are overall consistent with previous numerical studies similarly modeling such emission signatures \citep{Dessart10,Lovegrove13}. 

While we discuss the early peak in detail later in Section \ref{sec:lc_withCSM}, we can compare the plateau emission with previous studies. The plateau emission is brighter and shorter for larger explosion energies, due to earlier dilution and cooling of the envelope for faster ejecta velocities. \cite{Piro13} estimated the characteristic luminosity and duration of the plateau, by extrapolating the scalings found from numerical light curve models of Type II-P SNe \citep{Kasen09} to low energies, as\footnote{As noted in \cite{Sukhbold16} the equations in \cite{Kasen09} had typographic errors on the scaling of the helium mass fraction $X_{\rm He}$, which propagated in the equations of \cite{Piro13}. The scalings here are from the corrected equations in \cite{Sukhbold16}, with $X_{\rm He}\approx 0.33$ in our RSG models.}
\rev{\begin{eqnarray}
    t_{\rm pl} &\approx& 740\ {\rm days}\left(\frac{E_{\rm exp}}{10^{48}\ {\rm erg}}\right)^{\!\!-1/4} \!\! %\nonumber\\
   \left(\frac{M_{\rm ej}}{10~M_\odot}\right)^{\!\!1/2} \!\! \left(\frac{R_*}{800~R_\odot}\right)^{\!\!1/6}\\
    L_{\rm pl} &\approx& 5.5\times 10^{39}\ {\rm erg\ s^{-1}}\left(\frac{E_{\rm exp}}{10^{48}\ {\rm erg}}\right)^{\!\!5/6} \nonumber\\
    &&\times \left(\frac{M_{\rm ej}}{10~M_\odot}\right)^{\!\!-1/2} \left(\frac{R_*}{800~R_\odot}\right)^{\!\!2/3}.
\end{eqnarray}
For our $15~M_\odot$ ($20~M_\odot$) models with $R_*\approx 825~R_\odot$ ($1059~R_\odot$) and ejecta mass $M_{\rm ej}\approx$ 8.4$~M_\odot$ (8.6$~M_\odot$) almost independent of $E_{\rm exp}$, the above equations for $E_{\rm exp}=[10^{48}$, $3\times 10^{48}$, $10^{49}$] erg yield
\begin{eqnarray}
    t_{\rm pl} &\approx& 
    \begin{cases}
       [680, 520, 380]~{\rm days} & (15~M_\odot\ {\rm model}) \\
       [720, 550, 410]~{\rm days} & (20~M_\odot\ {\rm model}),
    \end{cases} \\
    L_{\rm pl} &\approx& 
    \begin{cases}
       [0.6, 1.5, 4.1]\times 10^{40}~{\rm erg\ s^{-1}} & (15~M_\odot\ {\rm model}) \\
       [0.7, 1.8, 4.8]\times 10^{40}~{\rm erg\ s^{-1}} & (20~M_\odot\ {\rm model}).
    \end{cases}
\end{eqnarray}}
While our results agree well with these equations within a factor of a few, it is worth pointing out two key differences in these low-energy explosions from SN II-Ps, that are also explosions of RSGs but with larger $E_{\rm exp}\sim 10^{51}$ erg. First, due to the high densities in the ejecta of 
\begin{eqnarray}
    \rho_{\rm ej}&\approx& \frac{M_{\rm ej}}{4\pi (t\sqrt{2E_{\rm exp}/M_{\rm ej}})^3/3} \nonumber \\
     &\!\! \sim& 2 \! \times \! 10^{-10}{\rm g\ cm^{-3}} \! \left(\frac{t}{{\rm yr}}\right)^{\!\!-3} \!\! \left(\frac{E_{\rm exp}}{10^{48}\ {\rm erg}}\right)^{\!\!-3/2} \! \! \left(\frac{M_{\rm ej}}{10\ M_\odot}\right)^{\!\!5/2} 
\end{eqnarray}
sustained during the plateau phase, the opacity in the ionized part of the ejecta $\kappa_{\rm ej}$ also remains higher than that expected from electron scattering, by up to an order of magnitude (Figure \ref{fig:kappa_OPAL}). This reduces the plateau luminosity, which scales with the characteristic opacity in the ionized ejecta as $\kappa_{\rm ej}^{-1/3}$ \citep{Popov93}, by up to a factor $\sim 2$.

Second, the recombination energy, released as the expanding hydrogen-rich envelope recombines, becomes increasingly important for lower-energy explosions. The energy budget from recombination is mostly independent of explosion energy and is evaluated as
\begin{eqnarray}
    E_{\rm recom} &=& (13.6\ {\rm eV})\frac{X_{\rm H}M_{\rm ej}}{m_p} \nonumber \\ 
    &\approx& 1.7\times 10^{47}\ {\rm erg}\left(\frac{X_{\rm H}}{0.65}\right)\left(\frac{M_{\rm ej}}{10~M_\odot}\right)
\end{eqnarray}
where $m_p$ is the proton mass and $X_{\rm H}$ is the hydrogen mass fraction of the ejecta. This energy is released near the recombination front beyond which the opacity sharply drops, and hence can be radiated efficiently without strong adiabatic losses. The radiated energy in the plateau obtained by integrating the light curve ranges as ($2$--$5)\times 10^{47}$ erg,  with lower values for lower $E_{\rm exp}$. This suggests that recombination power is a significant or even dominant contribution to the plateau.

Comparing the thin and thick lines in Figure \ref{fig:light_curves_noCSM}, the difference in light curves for the $15~M_\odot$ and $20~M_\odot$ models is small. For a fixed $E_{\rm exp}$, light curves of the $20~M_\odot$ model have slightly longer duration and higher luminosity in both breakout and plateau, due to the higher envelope mass and larger stellar radius.

We compare this plateau emission to a failed SN candidate N6946-BH1 \citep{Gerke15,Adams17a,Basinger21}, a RSG that displayed a months-long outburst before fading out in the optical. The outburst luminosity is estimated as $\sim 10^6$--$10^7~L_\odot$ for varying assumptions on extinction \citep{Adams17a}, consistent with the model light curves for $E_{\rm exp}\approx 10^{48}$--$10^{49}$ ergs. However the duration of the transient is constrained to $3$--$11$ months, somewhat shorter than the models. This may be explained by a lower ejecta mass than our models, due to enhanced envelope mass-loss of the RSG progenitor (as suggested in a sub-population of short-plateau SN II-Ps; \citealt{Hiramatsu21}). Another possibility is dust formation in the ejecta during the plateau phase, although this may only affect the light curves near the end of the plateau \citep{Kochanek14b}.

\subsection{Models with Dense CSM}
\label{sec:lc_withCSM}

As the CSM has a mass ($\sim 0.01$--$0.1~M_\odot$) much lower than the ejecta, the plateau emission is almost the same for models with and without dense CSM. Large differences appear in the early phase of the light curve around shock breakout. Hereafter we discuss the impacts of the CSM for the $15~M_\odot$ model (the $20~M_\odot$ model produces very similar results), with two explosion energies of $10^{48}$ and $10^{49}$ erg. 

\begin{figure*}
   \centering
    \begin{tabular}{cc}
     \begin{minipage}[t]{0.5\hsize}
    \centering
    \includegraphics[width=0.95\linewidth]{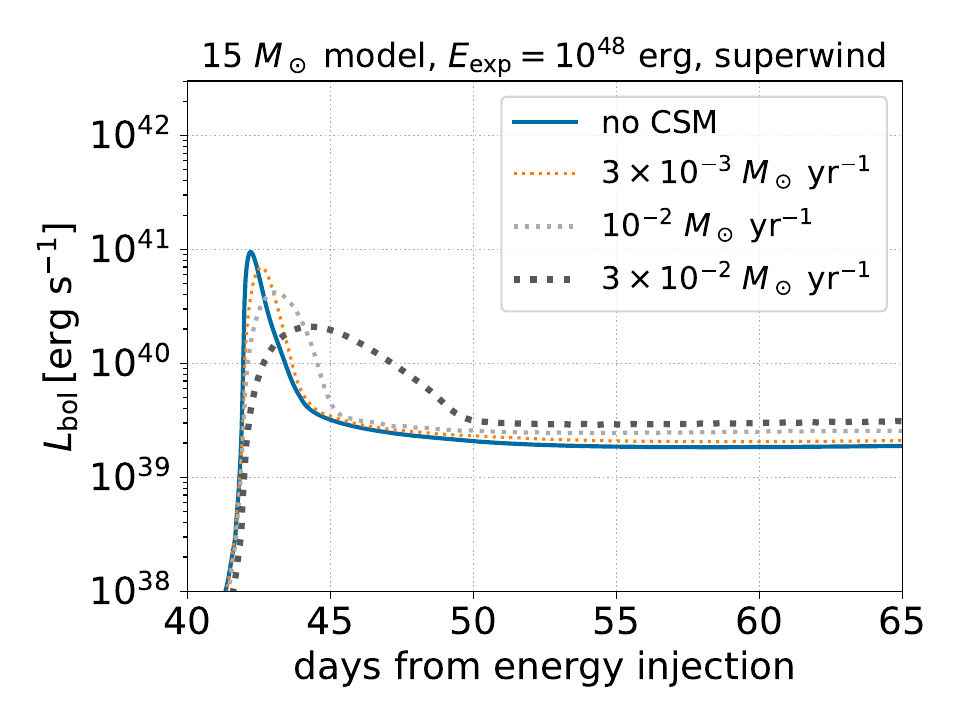}
    \end{minipage}
     \begin{minipage}[t]{0.5\hsize}
   \centering
    \includegraphics[width=0.95\linewidth]{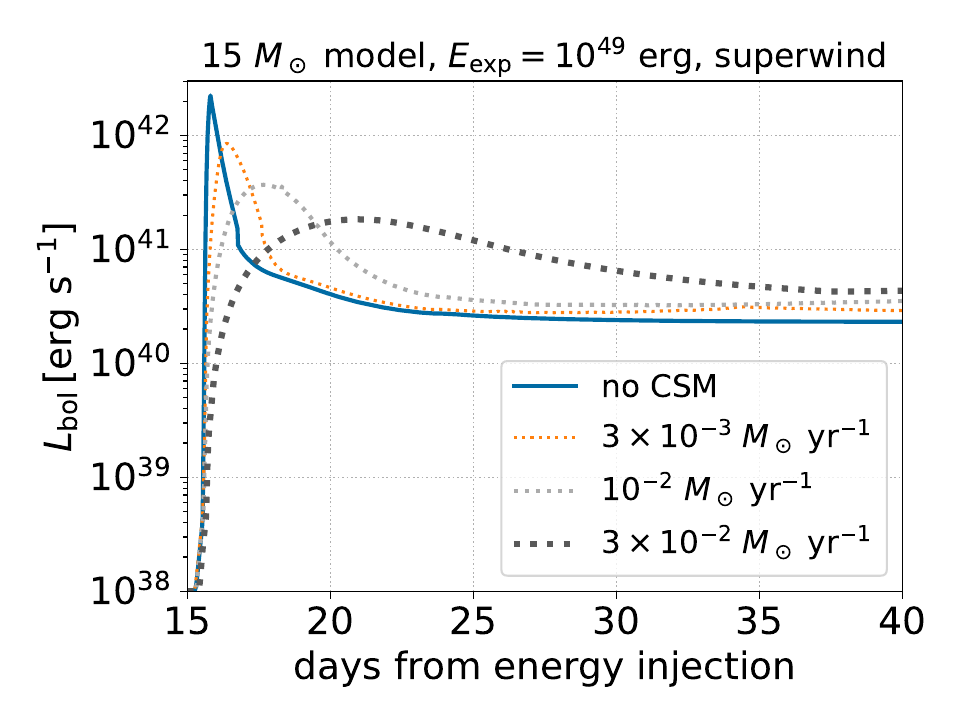}
    \end{minipage} \\
    \begin{minipage}[t]{0.5\hsize}
    \centering
    \includegraphics[width=0.95\linewidth]{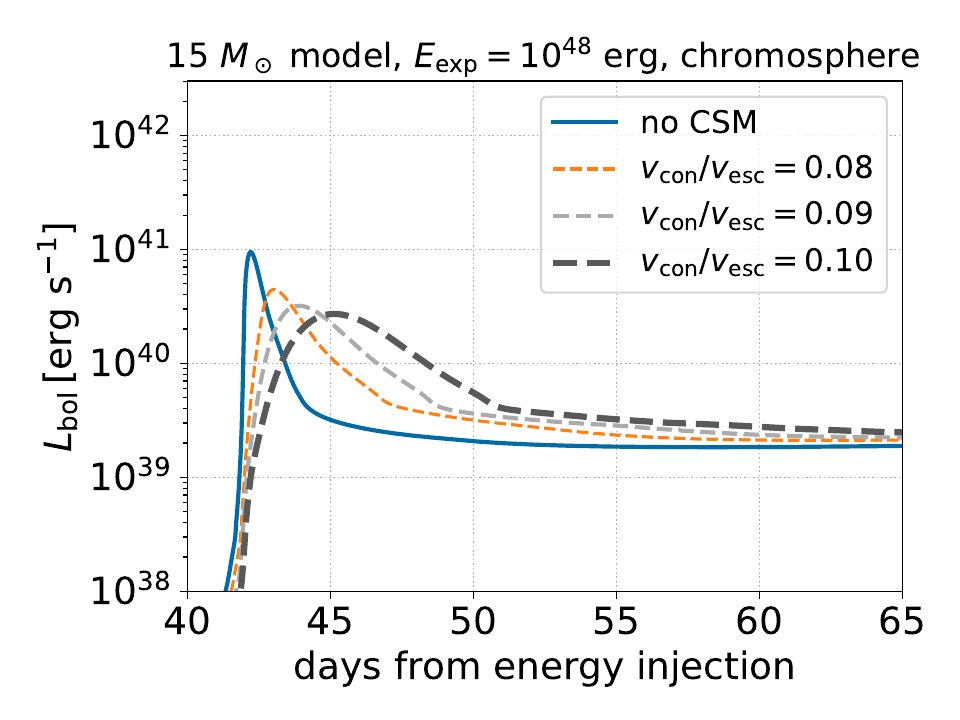}
    \end{minipage}
     \begin{minipage}[t]{0.5\hsize}
   \centering
    \includegraphics[width=0.95\linewidth]{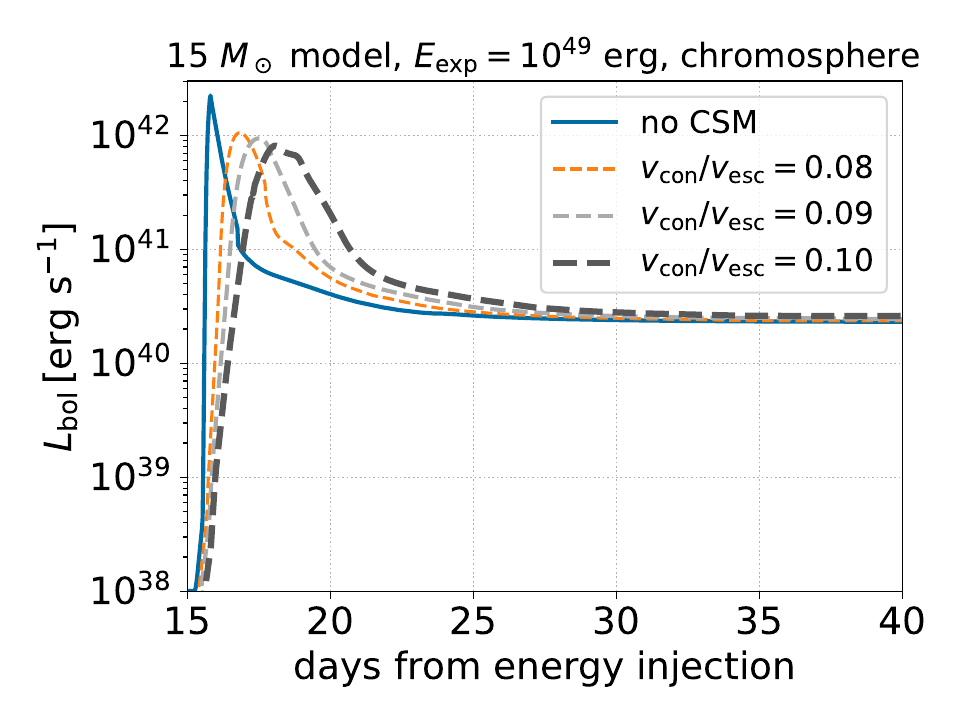}
    \end{minipage} 
    \end{tabular}
    \caption{Bolometric light curves including models with dense CSM, for a $15~M_\odot$ model with final explosion energies of $E_{\rm exp}=10^{48}$ (left panels) and $10^{49}$ erg (right panels). The top panels correspond to confined superwind models, and the bottom panels correspond to the dense chromosphere models. Solid lines are for no CSM, and dotted (dashed) lines are for superwind (chromosphere) models with varied mass loss rates ($v_{\rm con}/v_{\rm esc}$).}
    \label{fig:light_curves_denseCSM}
\end{figure*}

Figure \ref{fig:light_curves_denseCSM} shows the early light curves around shock breakout, for cases with and without ambient dense CSM. The general effect of adding a dense CSM is to make the breakout emission longer and dimmer, due to enhancement in the diffusion time in the CSM as it is ionized by the shock breakout emission. Once diffusion is complete, the light curves merge with the case without dense CSM, and transitions to the plateau phase.

For the chromosphere CSM models of \cite{Fuller24}, we find the timescale of the early peak to be $\approx 3$ -- 10 days, enhanced from $\lesssim 1$ day in the case of no dense CSM. The timescales are longer for larger $v_{\rm con}/v_{\rm esc}$ and lower $E_{\rm exp}$. The superwind CSM models inspired by \cite{Jacobson-Galan24} predict longer timescales for larger $\dot{M}$, also showing the effect of photon diffusion in the CSM. For the densest case of $\dot{M}=3\times 10^{-2}~M_\odot\ {\rm yr^{-1}}$ (with total CSM mass of $\approx 0.1~M_\odot$ within $5\times 10^{14}$ cm), the early peak becomes as long as weeks. The most energetic model of $10^{49}$ erg has additional non-negligible contribution from interaction with the extended superwind (with luminosity scaling as $L\propto (\dot{M}/v_{\rm wind})(v_{\rm sh}-v_{\rm wind})^3$), which also prolongs and flattens the decay. The continued interaction also slightly enhances the luminosity at the plateau phase, typically by $\lesssim$ (a few) $10\%$. This is an upper limit, as we assume stationary CSM in our simulations.

In some cases the two CSM models display similar light curves, despite large differences in the density profiles near the star. This is seen for example in the chromosphere model with $v_{\rm con}/v_{\rm esc}=0.1$ and superwind model with $\dot{M}=3\times 10^{-2}~M_\odot\ {\rm yr^{-1}}$, for $E_{\rm exp}=10^{48}$ erg. This indicates that bolometric light curves alone are insufficient if one aims to distinguish the CSM model from these transients, a degeneracy also seen in Type II SNe \citep[e.g.,][]{Dessart23}.

In Appendix \ref{sec:semiana_model}, we construct a useful semi-analytical model to calculate the timescale and (bolometric) luminosity of the early peak from these explosions, applicable for a wide range of CSM density profiles and also for bare RSGs without CSM. The model predicts values that agree well with our numerical results by SNEC.

\section{Prospects for Optical/UV Surveys}
\label{sec:discussion}

To consider prospects for near-future transient surveys by Rubin and ULTRASAT, we further analyze the outputs from SNEC and obtain the optical (620 nm) and near-UV (250 nm) light curves assuming thermal emission. Using the bolometric luminosity at the outer edge $r_{\rm edge}$ and photospheric radii $R_{\rm ph}(t)$ measured at an optical depth of $2/3$, we obtain the effective temperature
\begin{eqnarray}
    T_{\rm eff}(t) &=& \left(\frac{L(t)}{4\pi R^2_{\rm ph}(t'=t-t_{\rm lt})\sigma_{\rm SB}}\right)^{\!\!1/4} \nonumber \\
    &\sim& 1.1\times 10^4\ {\rm K}\left(\frac{L}{10^{41}\ {\rm erg\ s^{-1}}}\right)^{\!\!1/4}\left(\frac{R_{\rm ph}}{10^{14}\ {\rm cm}}\right)^{\!\!-1/2}
\end{eqnarray}
where $t_{\rm lt}\approx [r_{\rm edge}(t)-R_{\rm ph}(t)]/c$ is the time for light to radially travel from the photosphere to the outer edge\footnote{This is under the approximation that both the outermost (Lagrangian) cell at $r_{\rm edge}$ and the photosphere move much slower than $c$. The latter breaks down at the onset of shock breakout, when the first photons propagate through the cold CSM and the opacity in the CSM suddenly changes at a timescale comparable to $t_{\rm lt}$. It becomes valid just after shock breakout when the light curve starts to rise.}, and $\sigma_{\rm SB}$ is the Stefan-Boltzmann constant. We assume a blackbody spectra of radius approximated as $R_{\rm ph}$ (i.e.  temperature approximated as $T_{\rm eff}$)
\begin{eqnarray}
    L_\nu &=& 4\pi R_{\rm ph}^2\frac{2\rev{\pi} h\nu^3}{c^2}\frac{1}{\exp[h\nu /k_BT_{\rm eff}]-1},
\end{eqnarray}
where $h, k_B$ are respectively the Planck and Boltzmann constant. We then calculate the AB magnitude by
\begin{eqnarray}
    M_{\rm AB}= -2.5\log_{10}\left[\frac{L_\nu/4\pi (10~{\rm pc})^2}{3631\ {\rm Jy}}\right],
\end{eqnarray}

The assumption of thermal spectra is likely a good approximation for shock breakout from bare RSGs without dense CSM, due to the low shock velocity and high density $\sim 10^{-9}\ {\rm g\ cm^{-3}}$ at the breakout radius \citep{Piro13,Lovegrove17}. However, for models with dense CSM, the photosphere at around the bolometric peak is found to be within the CSM, at radii of $R_{\rm ph}\sim (1$--$3)\times 10^{14}$ cm with much lower densities of $\rho_{\rm ph}\sim (10^{-13}$--$10^{-12})\ {\rm g\ cm^{-3}}$. At these densities, the scattering opacity is comparable to or larger than the absorption opacity (Figure \ref{fig:kappa_OPAL}), and thermalization at the photosphere is not trivial. 

The required time for a gas of density $\rho_{\rm ph}$ to generate enough photons to realize a blackbody of temperature $T_{\rm eff}$ is \citep{Nakar10}
\begin{eqnarray}
    t_{\rm BB} \sim 3\times 10^3\ {\rm sec}\left(\frac{\rho_{\rm ph}}{10^{-13}\ {\rm g\ cm^{-3}}}\right)^{\!\!-2}\left(\frac{T_{\rm eff}}{10^4\ {\rm K}}\right)^{\!\!7/2}
\end{eqnarray}
Here we updated the photon production process from free-free to bound-free emission/absorption (appropriate for temperatures of $\sim 10^4$ K), which reduces the timescale by a factor of $\kappa_{\rm bf}/\kappa_{\rm ff}\sim 20$ for gas of solar metallicity (Section 2.2.5 of \citealt{Nakar10}). This is comparable to the light crossing time at the photosphere $R_{\rm ph}/c\sim 3\times 10^3\ {\rm sec}\ (R_{\rm ph}/10^{14}\ {\rm cm})$, so thermalization at the photosphere is only marginally justified at around peak. 
In future work we plan to further explore the light curves with multi-group radiation transport (as in e.g., \citealt{Lovegrove17}), to more accurately determine the spectral energy distribution at the early phases.

\begin{figure*}
   \centering
    \includegraphics[width=0.95\linewidth]{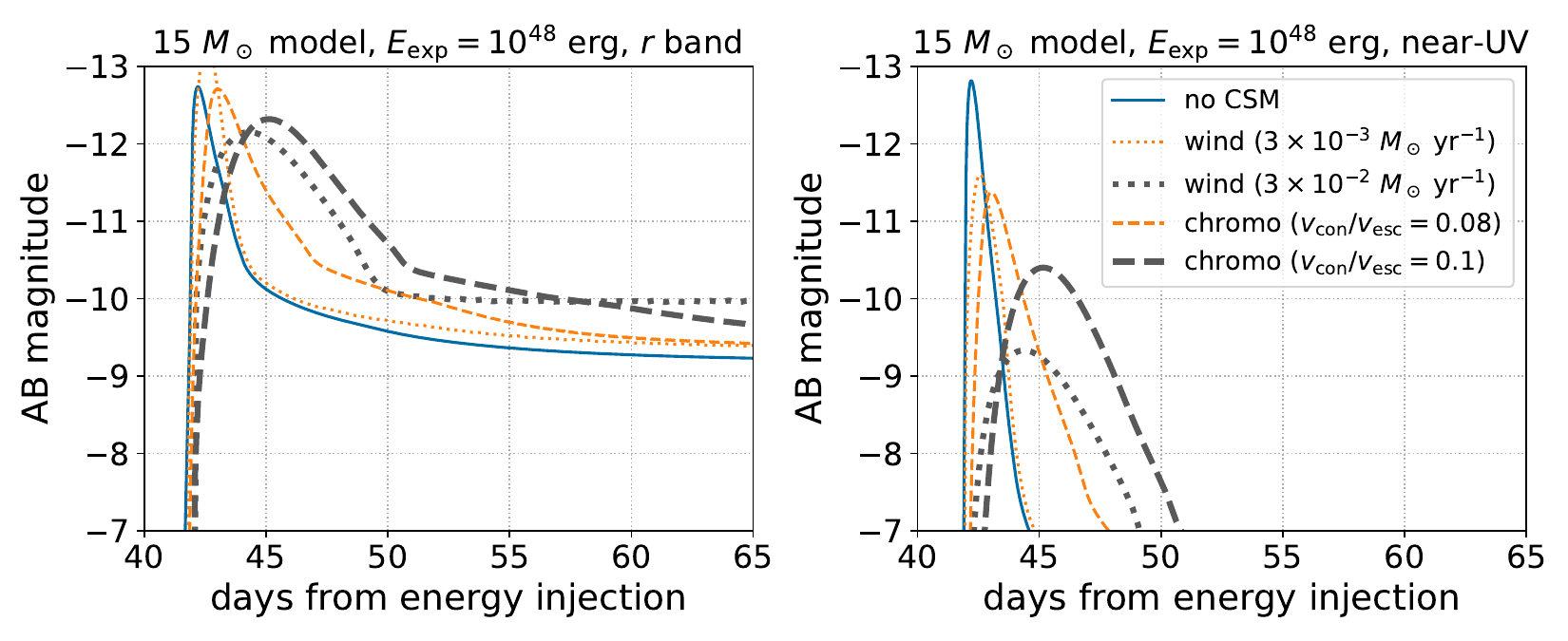} \\
    \includegraphics[width=0.95\linewidth]{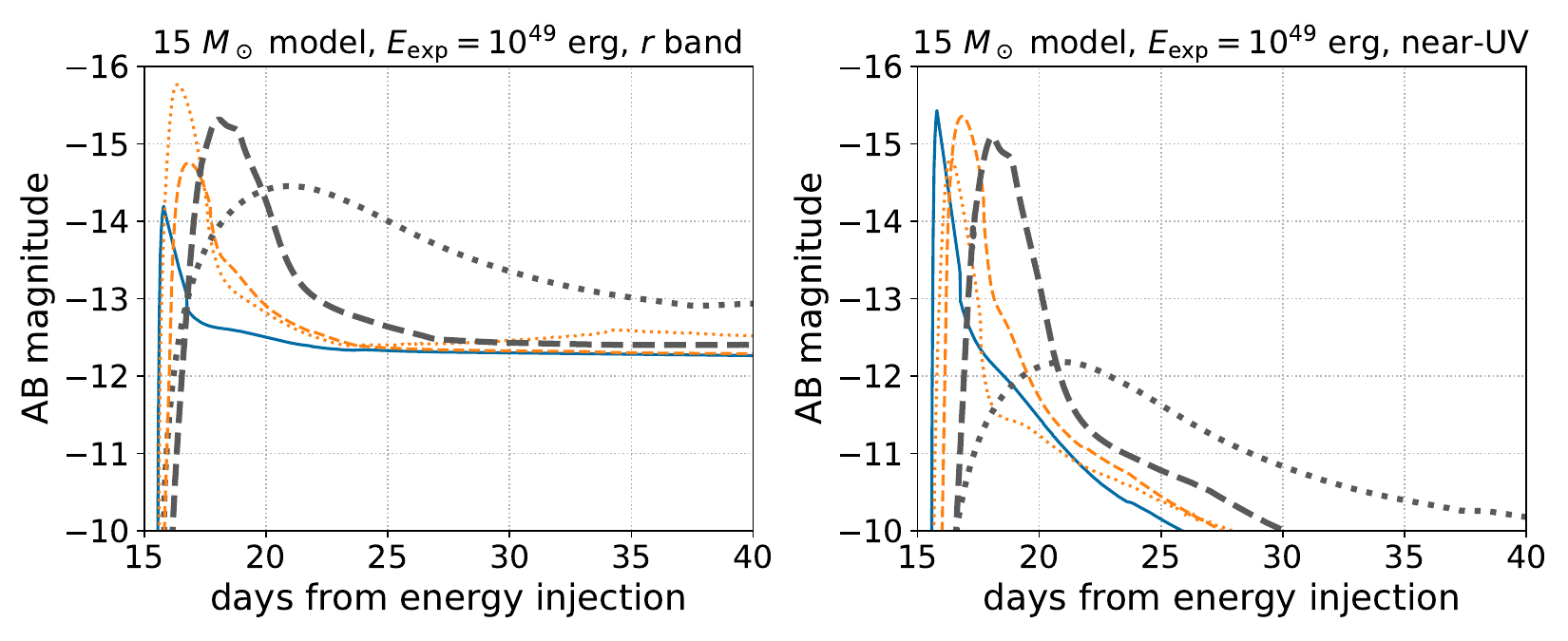}
    \caption{Absolute AB magnitudes for selected light curve models with $E_{\rm exp}=10^{48}$ erg (top) and $10^{49}$ erg (bottom), in the $r$ band (left) and near-UV band (right).}
    \label{fig:ABmag}
\end{figure*}

Figure \ref{fig:ABmag} shows the absolute AB magnitude for selected $15~M_\odot$ RSG models in the $r$ band (for Rubin) and near-UV (for ULTRASAT), respectively measured at a central wavelength of $620$ nm and $250$ nm. The predicted peak magnitudes in the optical and UV scale with the explosion energy, with typical ranges of \rev{($-9$)--($-12$)} mag, \rev{($-12$)--($-15$)} mag for energies of $E_{\rm exp}=10^{48}$ erg, $10^{49}$ erg respectively.

For Rubin, the fiducial cadence is 3 days but can be around 10 days when focused on an individual filter, with $r$ band having the highest cadence of 7 days \citep{Bianco22,Li22}. While the $r$ band depth of $24.7$ mag makes Rubin highly sensitive for early peaks out to $\gtrsim 100$ Mpc, their fast evolution may make them challenging for Rubin to observe in multiple epochs, except for the densest superwind CSM models with slowest evolution. However, for a fraction of the time ($\sim 10\%$; \citealt{Bellm22}) the observation gaps could be as short as $\approx 1$ day, and surveys of $1$ day cadence may be achieved for a subset of the deep drilling program \citep[e.g.,][]{Gris23}. For a failed SN rate of 20\% \citep{Neustadt21} of the core-collapse SN rate ($R_{\rm SN}\approx 10^{-4}~{\rm Mpc^{-3}\ {\rm yr}^{-1}}$; \citealt{Perley20}), we expect $40$ events per year within 100 Mpc and within a survey area of $18,000$ deg$^2$ covered by Rubin. Thus these subsets of short cadence observations may bear fruit on capturing the bright early peak.

% The non-detections of these failed SN events by PTF/ZTF have led to constraints on the failed SN rate \citep{Bryne22}, but such analysis focused on the year-long plateau. It may be worth conducting a similar search for the short-duration early peak, although the predicted light curves have larger variations which may preclude robust constraints for non-detections.

For ULTRASAT, the single-visit depth of 22.5 mag \citep{Shvartzvald24} would make it sensitive to more nearby failed SNe within $100$ Mpc, and only for the early shock breakout emission\footnote{UVEX, with a single-visit depth of 24.5 mag in UV \citep{Kulkarni21}, can detect sources out to a factor of a few farther than ULTRASAT. The detectability depends on the undetermined cadence and field of view of their high-cadence survey.}. We suggest that the high-cadence and low-cadence surveys proposed in ULTRASAT are complementary for failed SNe. The high-cadence survey with cadence of minutes would detect even the fastest evolving events, which enables early follow-up with other optical telescopes. The low-cadence survey, designed as single visits for a much wider field of view ($\approx 20\%$ of the sky), will instead more likely detect much closer events with detection horizon of $\approx$ a few $10$ Mpc. The expected event rate within $30~{\rm Mpc}$ is $\sim 0.2(0.2R_{\rm SN})[4\pi (30\ {\rm Mpc})^3]/3\approx 0.5$ events per year. Up to half of the RSG progenitors within this distance, in the observing regions overlapping with Rubin, are detectable via pre-explosion observations \citep{Strotjohann24}. These closer events are favorable for constraining the type of RSGs that do not explode as SNe.

We finally note that the early UV peak may be affected by the surviving dust in the pre-existing RSG wind. The prompt UV flash is expected to sublimate dust as it propagates outwards, out to a radii \citep{Waxman00,Lu16,Metzger23}
\begin{eqnarray}
    r_{\rm sub}\sim 1\times 10^{16}\ {\rm cm}\left(\frac{L_{\rm UV}}{10^{41}\ {\rm erg\ s^{-1}}}\right)^{\!\!1/2}\left(\frac{a_{\rm gr}}{0.1\ {\rm \mu m}}\right)^{\!\!-1/2}
\end{eqnarray}
where we adopt $a_{\rm gr}\sim 0.1~{\rm \mu}$m as the typical grain size, and a sublimation temperature of $1700\ {\rm K}$ for astronomical silicates. Any dust in the dense CSM would be sublimated, and surviving dust would be in the low-density wind emitted much earlier from the RSG, centuries before core-collapse. For a fiducial mass loss rate of $\dot{M}_{\rm RSG}=10^{-5}~M_\odot\ {\rm yr}^{-1}$ (which is around the upper end for observed RSGs; \citealt{Beasor20,Bietenholz21}), the UV optical depth by surviving dust at $r>r_{\rm sub}$ is
\begin{eqnarray}
    \tau_{\rm UV}&\approx& \frac{\kappa_{\rm UV}\dot{M}_{\rm RSG}}{4\pi v_{\rm w}r_{\rm sub}} \nonumber \\
    &\sim& 0.3\left(\frac{\dot{M}_{\rm RSG}}{10^{-5}\ M_\odot\ {\rm yr}^{-1}}\right) \!\! \left(\frac{v_w}{30\ {\rm km s}^{-1}}\right)^{\!\!-1} \!\! \left(\frac{a_{\rm gr}}{0.1\ {\rm \mu m}}\right)^{\!\!1/2}\nonumber \\
    &&\times \left(\frac{\kappa_{\rm UV}}{200\ {\rm cm^2\ g^{-1}}}\right)\left(\frac{L_{\rm UV}}{10^{41}\ {\rm erg\ s^{-1}}}\right)^{\!\!-1/2}
\end{eqnarray}
where we adopted a silicate dust UV opacity of $4\times 10^{4}\ {\rm cm^2\ g^{-1}}$ \citep[][]{Draine84,Inoue20} and a gas-to-dust ratio for solar metallicity of 200 \citep{vanloon05}. Therefore the UV predictions are generally safe for our models of $E_{\rm exp}\gtrsim 10^{48}$ erg that predict $L_{\rm UV}\gtrsim 10^{40}$ erg s$^{-1}$, but may be affected by dust extinction for much lower energies, or if extreme mass loss rates exceeding (a few) $\times 10^{-5}\ M_\odot\ {\rm yr}^{-1}$ are realized centuries before the SN.

\section{Conclusion}
\label{sec:conclusion}

Using a one-dimensional radiation hydrodynamical code SNEC, we have explored the light curves of weak explosions predicted in BH formation from RSGs. We have specifically focused on modeling cases where the dying RSGs have a confined dense CSM, which are commonly observed in Type II SNe arising from explosions of similar RSGs with energies of $\sim 10^{51}$ erg.

We found that the dense CSM prolongs the duration of the early bright peak (shock breakout) to $\sim$days-weeks, making them good targets for wide-field UV/optical surveys such as ULTRASAT and Rubin. The peak luminosities of $10^{7}$--$10^{8}~L_\odot$, an order of magnitude brighter than the later plateau and emitted mainly in UV and optical, are bright enough for these surveys to detect in the nearby universe where a significant fraction of their putative RSG progenitors can also be observable.

Our work is the first step towards incorporating realistic dense CSM in observational predictions of BH births, and this can be improved in various ways. In order to explore the broad parameter space expected for BH formation from RSGs, our calculations are limited to simulations assuming spherical symmetry. In reality, convective processes that govern the structure of RSGs lead to global asymmetries in the envelope and the CSM, the latter of which is observed in early phases of SN 2023ixf \citep[e.g.,][]{Vasylyev23,Smith23}. Asymmetries in the CSM near the star can be especially important for further prolonging the rise in the light curves \citep{Goldberg22b}. Future multi-dimensional radiation hydrodynamical studies will be neeeded to explore this effect in depth.

We have only investigated the photometric light curves, and spectral modeling is an important next step to further explore the observable effects of dense CSM. The shocks in these weak explosions can still have sufficiently high velocities ($v_{\rm sh}\gtrsim 200~$ km s$^{-1}$ for $E_{\rm exp}\gtrsim 10^{48}$ erg) to ionize the ambient CSM \citep{Shull79,Sutherland17}, and produce recombination lines of hydrogen (and possibly heavier metals) like canonical Type II SNe. While the slower shock velocities create a longer time window to observe the effects of dense CSM, the line emission itself can be much dimmer due to the smaller budget of ionizing photons. In future work, we plan to conduct multi-group simulations and spectral modeling, more suitable for comparison with observational studies of these unique events in both optical and UV.

\section*{acknowledgments}

We thank Re'em Sari and Takashi Moriya for valuable discussions, and the anonymous referee for the constructive comments. D. T. and X. H. are supported by the Sherman Fairchild Postdoctoral Fellowship at the California Institute of Technology. The light curve data is publicly available in \url{https://github.com/DTsuna/failedSNeLCs}.

%\end*{acknowledgments}

\begin{appendix}
    
\section{Semi-analytical Breakout Estimates}
\label{sec:semiana_model}

We build a simple semi-analytical framework for modeling the effect of dense CSM on the early breakout peak. This can be used to roughly estimate the characteristic quantities of the early breakout emission (timescale, luminosity) without numerical simulations, and can be generalized to a wide range of envelope/CSM profiles.

\begin{figure}
    \centering
    \includegraphics[width=\linewidth]{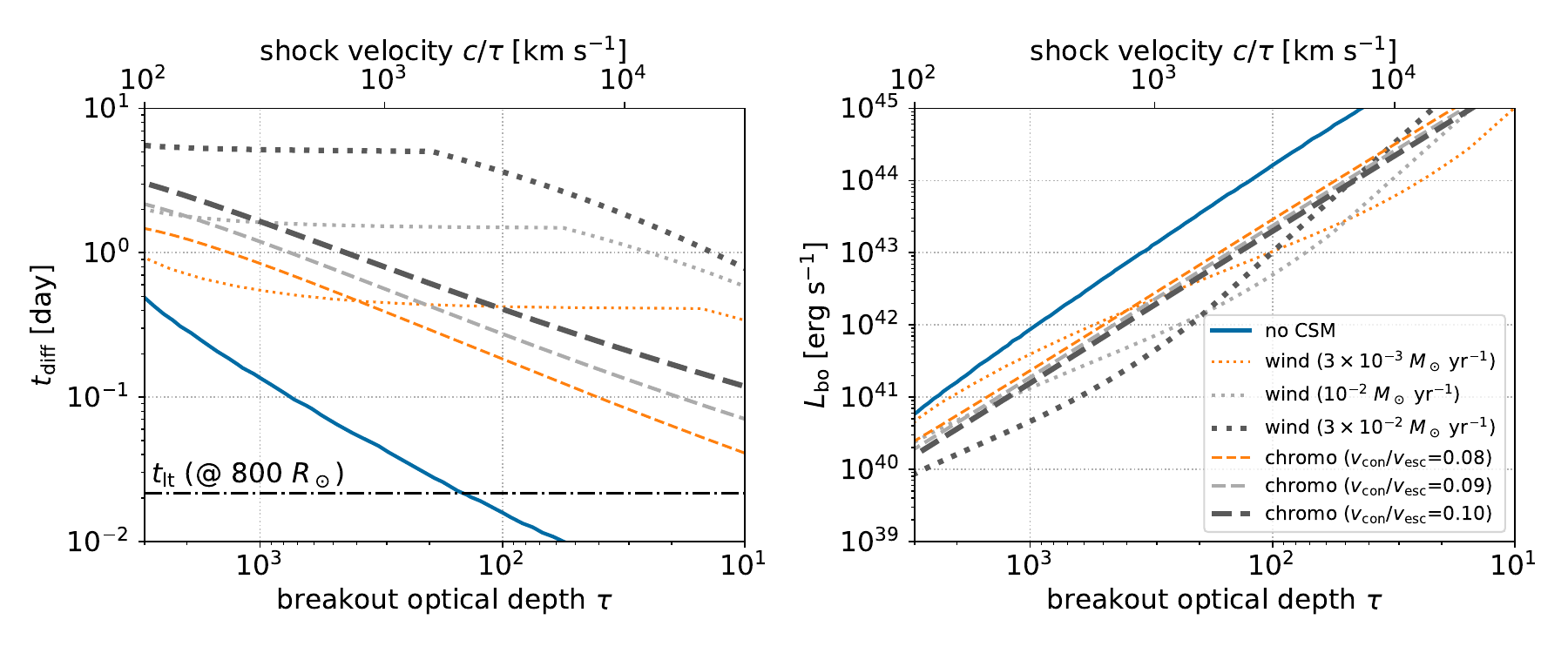}
    \caption{The expected timescale and luminosity of the breakout emission as a function of breakout optical depth $\tau$ (or shock velocity $c/\tau$), for a base RSG (solid line) and RSGs with CSM used in our simulations (dashed, dotted lines; see also Figure \ref{fig:CSM_profile}). The horizontal dash-dotted line on the left panel shows the light-travel time at a radius of $800~R_\odot$, roughly the stellar radius for the $15~M_\odot$ model.} %As illustrated, we have adopted the density profiles for the $15~M_\odot$ RSG model in our work (Figure \ref{fig:CSM_profile}). The horizontal dash-dotted line on the left panel shows the light-travel time at a radius of $800~R_\odot$, roughly the stellar radius for the $15~M_\odot$ model.}
    \label{fig:tdiff}
\end{figure}

Let us consider a shock with velocity $v_{\rm sh}$ propagating into an envelope/CSM with density profile $\rho(r)$. Shock breakout occurs at a radius $r_{\rm bo}$ where the optical depth ahead of the shock is
\begin{eqnarray}
\tau(r_{\rm bo})\approx \int_{r_{\rm bo}}^\infty (\kappa\rho) dr \approx \frac{c}{v_{\rm sh}}.    
\end{eqnarray}
The timescale of the observed breakout emission is set by the diffusion time of matter ahead of $r_{\rm bo}$ after it becomes ionized by the leaking photons. This can be evaluated as
\begin{eqnarray}
    t_{\rm diff}(r_{\rm bo})\approx \int_{r_{\rm bo}}^\infty \frac{dr}{c}(\kappa\rho H_\rho)
\end{eqnarray}
where $H_\rho \equiv \rho/|d\rho/dr|$ is the density scale height. The emission is further smeared over the light-travel time $t_{\rm lt}\approx r_{\rm bo}/c$, which is sub-dominant for weak explosions \citep[][see also Figure \ref{fig:tdiff}]{Piro13}. Note that for breakout deep within an extended wind ($\rho=Dr^{-2}$ where $D$ is constant) with constant opacity, this becomes a constant $t_{\rm diff}\sim \kappa D/c$ (with logarithmic dependence on the outer edge of the dense wind $r_{\rm out}$, or the photospheric radius) as obtained in other works \citep[e.g.,][]{Chevalier11,Ginzburg12,Haynie21}.

We next estimate the luminosity of the breakout pulse, by calculating the released radiation energy due to shock dissipation. The post-shock pressure is from the jump condition
\begin{eqnarray}
    p = \frac{2}{\gamma+1}\rho v_{\rm sh}^2 \sim 10^5\ {\rm erg\ cm^{-3}}\left( \frac{2}{\gamma+1}\right)\left(\frac{\rho}{10^{-9}\ {\rm cm^{-3}}}\right) \left(\frac{v_{\rm sh}}{100\ {\rm km\ s^{-1}}}\right)^{\!\!2}
\end{eqnarray}
where $\gamma$ is the adiabatic index. Whether radiation or gas dominates this depends on the shock velocity, with gas pressure dominating at low velocities \citep{Piro13}. We can include both regimes by solving for the gas-radiation equilibrium temperature under which either of the two dominates $p$. For radiation-dominated case 
\begin{eqnarray}
T_{\rm eq, rad}\approx \left(\frac{18\rho v_{\rm sh}^2}{7a}\right)^{\!\!1/4} \sim 7.6\times 10^4\ {\rm K}\left(\frac{\rho}{10^{-9}\ {\rm cm^{-3}}}\right)^{\!\!1/4} \left(\frac{v_{\rm sh}}{100\ {\rm km\ s^{-1}}}\right)^{\!\!1/2},  
\end{eqnarray}
where $a$ is the radiation constant, and in the gas-dominated case
\begin{eqnarray}
T_{\rm eq, gas}\approx \frac{3\mu m_pv_{\rm sh}^2}{16k_B}\sim 1.4\times 10^5\ {\rm K}\left(\frac{\mu}{0.62}\right) \left(\frac{v_{\rm sh}}{100\ {\rm km\ s^{-1}}}\right)^{\!\!2}.
\end{eqnarray}
where $k_B, m_p, \mu$ are respectively the Boltzmann constant, proton mass and mean molecular weight. We set the equilibrium temperature as $T_{\rm eq}\approx {\rm min}(T_{\rm eq, rad}, T_{\rm eq, gas})$, and obtain the fraction of internal energy contained in radiation
\begin{eqnarray}
    f_{\rm rad}(\rho, v_{\rm sh}) \approx \frac{aT_{\rm eq}^4}{aT_{\rm eq}^4 + 3/2(\rho/\mu m_p)k_B T_{\rm eq}}. 
\end{eqnarray}

Upon breakout, the radiation energy generated from shock dissipation over a diffusion time is released over the diffusion time, which results in a characteristic breakout luminosity
\begin{eqnarray}
    L_{\rm bo} \approx \frac{1}{t_{\rm diff}}\int_{r_{\rm bo}}^{r_{\rm bo}+v_{\rm sh}t_{\rm diff}}4\pi r^2 \rho v_{\rm sh}^2 f_{\rm rad} dr,
\end{eqnarray}
where we include the radial dependence of $\rho, f_{\rm rad}$ but neglect that of $v_{\rm sh}$ when conducting the integration.

A new element we add from previous analytical modeling \citep[e.g.,][]{Piro13} is the density dependence on the opacity $\kappa(\rho)$. As noted previously, for weak explosions the density at the breakout region ($\sim 10^{-9}\ {\rm g\ cm^{-3}}$) is high enough and the temperature low enough such that opacity can be greatly increased above the Thomson scattering value. While the opacity has a complicated temperature dependence which is captured in the SNEC calculations, for this semi-analytical approach we neglect this and adopt a dependence on the density as
\begin{eqnarray}
    \kappa(\rho) \approx 0.3\ {\rm cm^2\ g^{-1}}\left[1+\left(\frac{\rho}{10^{-11}\ {\rm g\ cm^{-3}}}\right)^{\!\!0.5}\right] .
    \label{eq:kappa_vs_rho}
\end{eqnarray}
The scaling with density in the second term in the square brackets is motivated from the neutral hydrogen opacity $\kappa_{\rm H^-}\approx 0.3\ {\rm cm^2\ g^{-1}}\ (\rho/10^{-11}\ {\rm g\ cm^{-3}})^{0.5}$, evaluated at $10^4$ K. While this prescription is very crude, it captures the density dependence of the OPAL opacities at $\rho\approx 10^{-12}$--$10^{-9}\ {\rm g\ cm^{-3}}$ within a factor of a few, at temperatures $10^4$--$10^5$ K relevant for weak explosions (see Figure \ref{fig:kappa_OPAL}). The model may be improved by self-consistently solving the temperature profile via steady-state radiation transport calculations for a given input luminosity \citep[e.g.,][]{Piro20}, using a temperature-dependent opacity instead of equation (\ref{eq:kappa_vs_rho}). However this approach requires knowing $L_{\rm bo}$ a priori, and hence is more complicated as it requires simultaneously solving for both $L_{\rm bo}$ and $t_{\rm diff}$ by iteration \citep[see][for a similar attempt]{Tsuna24}.

In Figure \ref{fig:tdiff} we show the timescale and luminosity of the breakout emission, $t_{\rm diff}$ and $L_{\rm bo}$, as a function of $\tau$ at breakout and the corresponding shock velocity $c/\tau$. For illustration we considered the initial density profiles in Section \ref{sec:CSM_models} used for our SNEC modeling. Generally the CSM increases the timescale and reduces the luminosity, as found from the SNEC simulations. The wind profile predicts an interesting dependence of $L_{\rm bo}$ on $\dot{M}$, with low $\dot{M}$ being brightest at low $v_{\rm sh}$ due to the shorter diffusion times, and high $\dot{M}$ being brightest at high $v_{\rm sh}$ (like those in successful SNe) due to the larger radiation energy budget gained from CSM interaction.

The failed SN models with $E_{\rm exp}=10^{48}$--$10^{49}$ ergs typically have $v_{\rm sh}\approx 100$--$400$ km s$^{-1}$ upon breakout and covers the left end of the plot, while successful SNe ($v_{\rm sh}\approx 3000$--$10^4$ km s$^{-1}$) covers the right half of the plot. For failed SNe, the predicted timescales and luminosities of the early peak from Figure \ref{fig:tdiff} are ($\sim 0.1$--$0.4$ day, $10^{41}$--$2\times 10^{42}$ erg s$^{-1}$) for models with no CSM, (1--10 days, $10^{40}$--$10^{42}$ erg s$^{-1}$) for the superwind models, and (1--3 days, $2\times 10^{40}$--$5\times 10^{41}$ erg s$^{-1}$) for the dense chromosphere models. These values reasonably agree with the early-phase light curves from the SNEC simulations.

\end{appendix}

\bibliography{references}

\end{document}